\documentclass[%
reprint,
superscriptaddress,
 amsmath,amssymb,
 aps,
pra,
]{revtex4-1}

\usepackage{graphicx}
\usepackage{dcolumn}
\usepackage{bm}

\usepackage{amsmath,amssymb}
\usepackage[capitalize]{cleveref}
\usepackage[separate-uncertainty=true,multi-part-units=single]{siunitx}
\usepackage{float}
\DeclareSIUnit\mbar{\milli\bar}
\providecommand{\config}[3]{\ensuremath{{#1}{#2}^{#3}}}
\providecommand{\term}[3]{\ensuremath{^{#1}\text{#2}_{#3}}}
\providecommand{\ani}[1]{\ensuremath{\text{#1}^{-}}}

\providecommand{\clu}[2]{\ensuremath{\text{#1}_{#2}^{-}}}

\providecommand{\cluH}[2]{\ensuremath{\text{#1}_{#2}\text{H}^{-}}}
\usepackage{xparse}
\usepackage{color}

\begin{document}

\title{Decays of excited silver cluster anions \clu{Ag}{n}, $n = 4$~to~$7$, in DESIREE}

\author{E.~K.~Anderson}
\email{emma.anderson@fysik.su.se}
\affiliation{Department of Physics, Stockholm University, AlbaNova, SE-106 91 Stockholm, Sweden}

\author{M.~Kami\'{n}ska}
\affiliation{Department of Physics, Stockholm University, AlbaNova, SE-106 91 Stockholm, Sweden}
\affiliation{Institute of Physics, Jan Kochanowski University, 25-369 Kielce, Poland}

\author{K.~C.~Chartkunchand}
\affiliation{Department of Physics, Stockholm University, AlbaNova, SE-106 91 Stockholm, Sweden}

\author{G.~Eklund}
\affiliation{Department of Physics, Stockholm University, AlbaNova, SE-106 91 Stockholm, Sweden}

\author{M.~Gatchell}
\affiliation{Department of Physics, Stockholm University, AlbaNova, SE-106 91 Stockholm, Sweden}
\affiliation{Institute for Ion Physics and Applied Physics, University of Innsbruck, Technikerstr.~25, A-6020 Innsbruck, Austria}

\author{K.~Hansen}
\affiliation{Center for Joint Quantum Studies and 
Department of Physics, Tianjin University, 92 Weijin Road, Tianjin 300072, China}
\affiliation{Department of Physics, University of Gothenburg, 41296 Gothenburg, Sweden}

\author{H.~Zettergren}
\affiliation{Department of Physics, Stockholm University, AlbaNova, SE-106 91 Stockholm, Sweden}

\author{H.~Cederquist}
\affiliation{Department of Physics, Stockholm University, AlbaNova, SE-106 91 Stockholm, Sweden}

\author{H.~T.~ Schmidt}
\affiliation{Department of Physics, Stockholm University, AlbaNova, SE-106 91 Stockholm, Sweden}

\date{\today}

\begin{abstract}

Spontaneous decays of small, hot silver cluster anions \clu{Ag}{n}~$n=4-7$ have been studied using one of the rings of the Double ElectroStatic Ion Ring ExpEriment (DESIREE). Observation of these decays over very long time scales is possible due to the very low residual gas pressure ($\sim10^{-14}$) and cryogenic (13 K) operation of DESIREE. The yield of neutral particles from stored beams of \clu{Ag}{6}~and~\clu{Ag}{7} anions were measured for 100 milliseconds and were found to follow single power law behaviour with millisecond time scale exponential cut-offs. The \clu{Ag}{4}~and~\clu{Ag}{5} anions were stored for 60 seconds and the observed decays show two-component power law behaviors. We present calculations of the rate constants for electron detachment from, and fragmentation of \clu{Ag}{4}~and~\clu{Ag}{5}. In these calculations, we assume that the internal energy distribution of the clusters are flat and with this we reproduce the early steep parts of the experimentally measured decay curves for \clu{Ag}{4} and \clu{Ag}{5}, which extends to tens and hundreds of milliseconds, respectively. The fact that the calculations reproduce the early slopes of \clu{Ag}{4} and \clu{Ag}{5}, which differ for the two cases, suggests that it is the changes in fragmentation rates with internal cluster energies of \clu{Ag}{4} and \clu{Ag}{5} rather than conditions in the ion source that determines this behavior. Comparisons with the measurements strongly suggest that the neutral particles detected in these time domains originate from \clu{Ag}{4}$\rightarrow$\clu{Ag}{3}$+$ Ag and \clu{Ag}{5}$\rightarrow$\clu{Ag}{3}$+$ Ag$_{2}$ fragmentation processes. 

\end{abstract}

\maketitle

\section{Introduction}

Access to mass-selected clusters of different sizes allow the study of how physical and chemical properties change when going from small systems with just a few atoms to systems with tens, hundreds or thousands of atoms, and to bulk matter. Small clusters are particularly interesting to study in this context as their various properties are expected to depend strongly on the number of atoms they contain. Here, spontaneous and/or photo-induced decays of various sorts may be used to probe the energetics and structural properties of these systems. Internal excitation energies in, for example, charged metal clusters can dissipate via different types of processes, including radiative cooling. Often several decay channels are open and may contribute to the decay on different time scales. Observations of electron detachment and/or fragmentation can then reveal features of photo-emission processes. 

The development of cryogenic electrostatic ion beam storage devices~\cite{Thomas2011, Schmidt2013, Reinhed2009, Lange2010, Hahn2011, Nakano2017} has made it possible to store ions for extended periods of time and to investigate relaxation processes on time scales ranging from microseconds to hours. Electrostatic ion storage devices have no upper mass limit, which facilitates the storage of heavy keV ion beams as demonstrated by the first electrostatic storage ring for atomic, molecular, and cluster physics -- the ELISA ring in Aarhus~\cite{Moller1997}. ELISA is operated at room temperature with a residual gas pressure of 10$^{-10} - 10^{-11}$ mbar (molecular number density $\sim10^{6}$~cm$^{-3}$), yielding typical ion beam storage times on the order of seconds. Cryogenic operation vastly improves ion beam storage capabilities as it lowers the residual gas pressure by orders of magnitude. Number densities between 10$^{2}$ and 10$^{4}$ per cm$^{3}$~\cite{Hahn2016, Lange2010, Schmidt2013} have been reported to give 1/e ion beam storage lifetimes of minutes and up to almost an hour~\cite{Lange2010, Hahn2016, Backstrom2015}. Equally important, cryogenic operation may allow the stored ions to approach thermal equilibrium with the device at temperatures down to a few kelvin~\cite{Thomas2011, Schmidt2013, Lange2010, Hahn2011, Nakano2017, Schmidt2017}.

In 2001, Hansen {\it et al.}~\cite{Hansen2001} reported on the spontaneous decay of internally hot metal cluster anions in ELISA. It was found that the production rate of neutrals leaving the ring varied with time $t$ after injection as $t^{-1+\delta}$ where $|\delta|<1$. This power law decay is caused by a broad distribution of decay constants, most often produced by a close-to-uniform distribution of internal excitation energies of the stored ions~\cite{Hansen2001}. Since then, power law decays have been reported for many different types of charged polyatomic systems including metal clusters~\cite{Kafle2015,Toker2007, Breitenfeldt2016, Fedor2005, Hansen2017}, fullerenes~\cite{Andersen2001,Tomita2006}, small carbon and hydrocarbon molecules~\cite{Goto2013, Shiromaru2015, Najafian2014}, biomolecules~\cite{Andersen2003_b, Andersen2004, Nielsen2004}, Polycyclic Aromatic Hydrocarbons (PAHs)~\cite{Martin2013, Ji2013}, and \clu{SF}{6}~\cite{Rajput2008, Menk2014}. In many of these cases, deviations from a pure $t^{-1}$ decay have been observed. These deviations are most frequently observed for systems consisting of a few atoms only and are discussed in terms of their specific properties such as heat capacities~\cite{Goto2013}, available decay channels~\cite{Aviv2011}, densities of final states~\cite{Menk2014} and the shape of the internal energy distribution~\cite{Ji2013}.

In a pioneering study, Hansen {\it et al.}~\cite{Hansen2001} reported power law decay behavior extending to milliseconds for \clu{Ag}{n} ($n=$\,4\,--\,9), with radiative cooling being significant on time scales of tens of milliseconds for clusters with $n~=$~4, 6, and 8. The \clu{Ag}{5} anion, however, followed a power law in their whole measurement window of \SI{50}{\ms}~\cite{Hansen2001}.

In this paper we present experimental studies of the spontaneous decay of small, internally hot silver cluster anions (\clu{Ag}{n}, $n =$~4 -- 7) as a function of storage time for up to 60 seconds in one of the DESIREE ion beam storage rings. Furthermore, we present calculations of vibrational autodetachment and fragmentation rates for the \clu{Ag}{4} and \clu{Ag}{5} clusters and discuss these in light of the present experimental results.

\section{Experimental apparatus and methods}

\begin{figure}
\centering
\includegraphics[width=1\columnwidth]{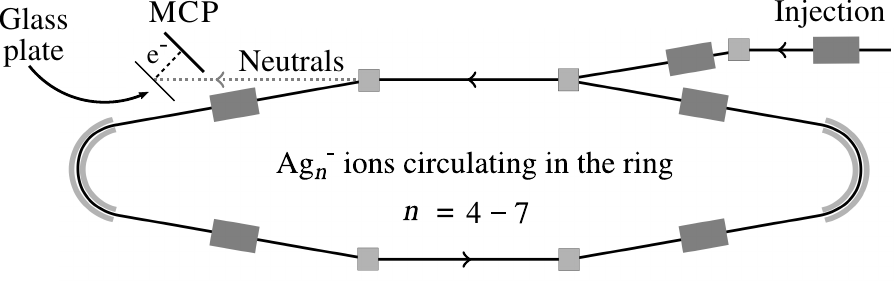}

\caption{A schematic of one of the DESIREE ion storage rings. Size-selected silver cluster anions, \clu{Ag}{n}, $n =$\,4\,--\,7, were injected and stored in the ring at 10\,keV. The neutral particles from spontaneous decay (fragmentation and/or electron detachment) of the excited clusters were counted by a detector system consisting of a glass plate with a gold-titanium film emitting secondary electrons and a micro-channel plate (MCP) detector.}

\label{fig:apparatus}
\end{figure}
 
A detailed description of the DESIREE ion beam storage rings can be found elsewhere~\cite{Thomas2011,Schmidt2013,Backstrom2015}, so only an outline of the salient features will be given here. Silver anions were produced using a Source of Negative Ions by Cesium Sputtering (SNICS~II)~\cite{SNICS} with a silver cathode. This type of ion source is known to produce a range of cluster sizes with broad internal energy distributions~\cite{Wucher1996}. The \clu{Ag}{n} ions were accelerated to \SI{10}{\keV} and mass-selected using a 90$^{\circ}$ analyzing magnet and injected into one of the ion storage rings of DESIREE, as shown in Fig.~\ref{fig:apparatus}. A chopped ion bunch of tens of microseconds duration, filling up half the ring upon injection, was stored and neutral particles were counted by means of a detector mounted along the line-of-sight of one of the straight sections of the ring. The detector assembly consists of a gold-titanium coated glass plate, a triple-stack micro-channel plate (MCP) and a resistive anode encoder (RAE). When neutral keV  particles impinge on the glass plate, secondary electrons are emitted and accelerated towards the MCP for detection as indicated in Fig.~\ref{fig:apparatus}.

Two sources of background need to be considered: detector dark counts and counts generated by neutrals due to collisions with the residual gas~\cite{Schmidt2013}. The detector dark count rate was measured between injections, with no ions in the ring, and subtracted from the rates measured after ion injection. Due to the extremely low residual gas density, the signal from residual gas collisions gives only a very small contribution in relation to the high neutral rates from spontaneously decaying hot clusters and is only visible after the initial decays discussed here have occurred, after seconds of storage or longer.

\begin{figure}
 \centering
 \includegraphics[width=1\columnwidth]{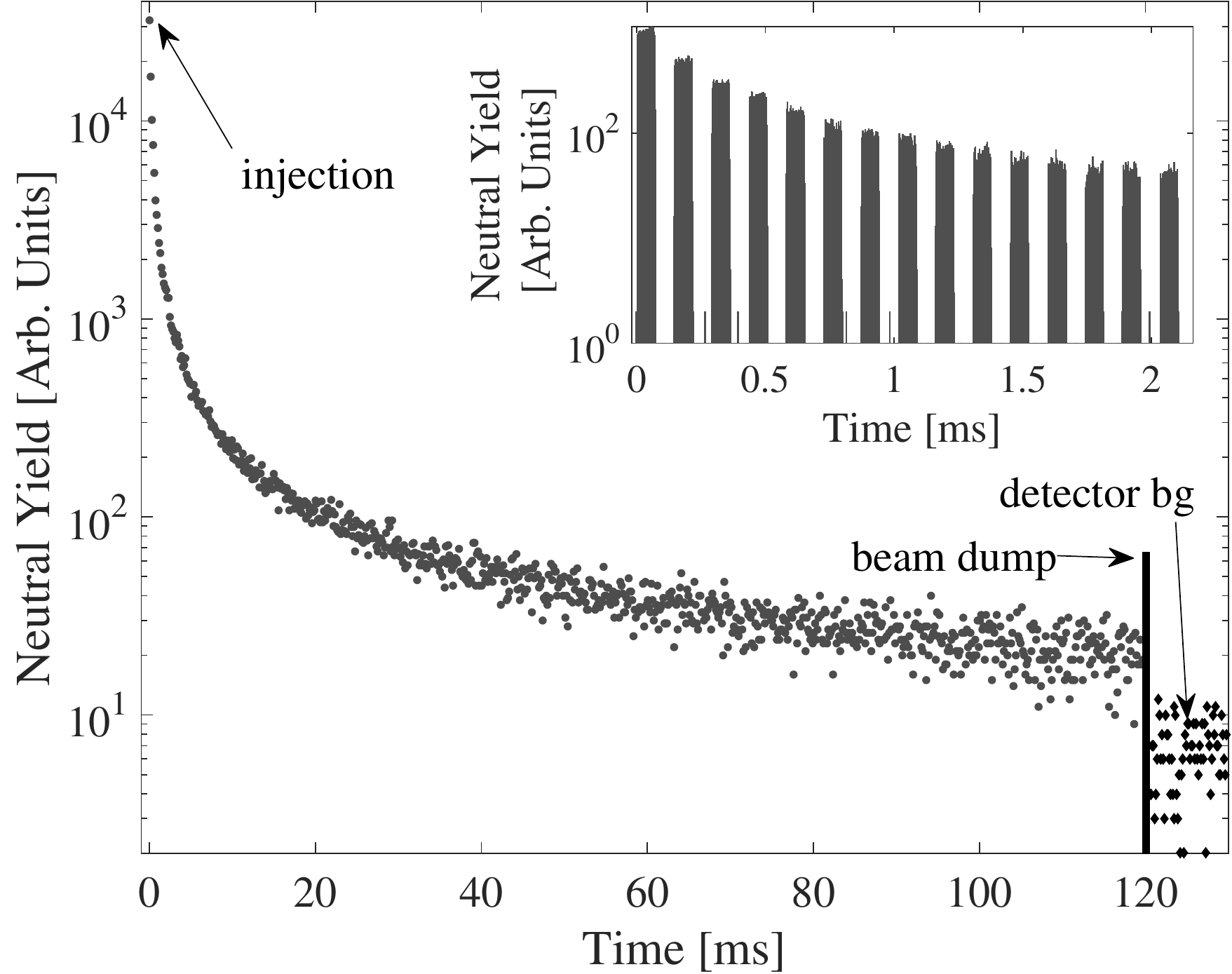}
 \caption {Yield of neutrals produced by a 10 keV \clu{Ag}{5} beam recorded as a function of time after injection. The inset shows the raw data for the first \SI{2}{\ms} (the first 15 turns can be seen) where the length of a single ion bunch corresponds to half a turn in the ring at injection. In the main figure, each point on the curve is a sum of the counts within a single turn. For this particular data set the measurement time window was chosen to be \SI{120}{ms}. The removal of the remaining ions (the beam dump), and the detector dark counts are indicated.}
 	 	
 \label{fig:bunches}
 \end{figure}

Measurements of neutral yields due to the stored clusters were performed by accumulating data over many ion-injection-and-storage cycles. Each measurement cycle started with injection of a mass-selected ion bunch, followed by storage of this beam for a pre-set time window, removal (dumping) of the stored beam, and measurement of the detector dark count background rate as shown in the example in Fig.~\ref{fig:bunches}. The storage time windows ranged from \SI{100}{\ms} to \SI{60}{\s} depending on the cluster ion.

\subsection {Ion beam storage capability as measured with \ani{Ag}} \label{beam storage}
 
The silver monomer anion, \ani{Ag}, has only one bound state (\config{4}{d}{10}\config{5}{s}{2} \term{1}{S}{0}) with a binding energy of 1.30447(2)\,eV~\cite{Bilodeau1998}. We used a laser-probing technique to measure the storage lifetime of the \ani{Ag} beam to gauge the general storage conditions for ions including the cluster ions of interest here. These measurements show that the ion storage time is orders of magnitude longer than the typical cluster decay times we are investigating here.

\begin{figure}
 \centering
 \includegraphics[width=1\columnwidth]{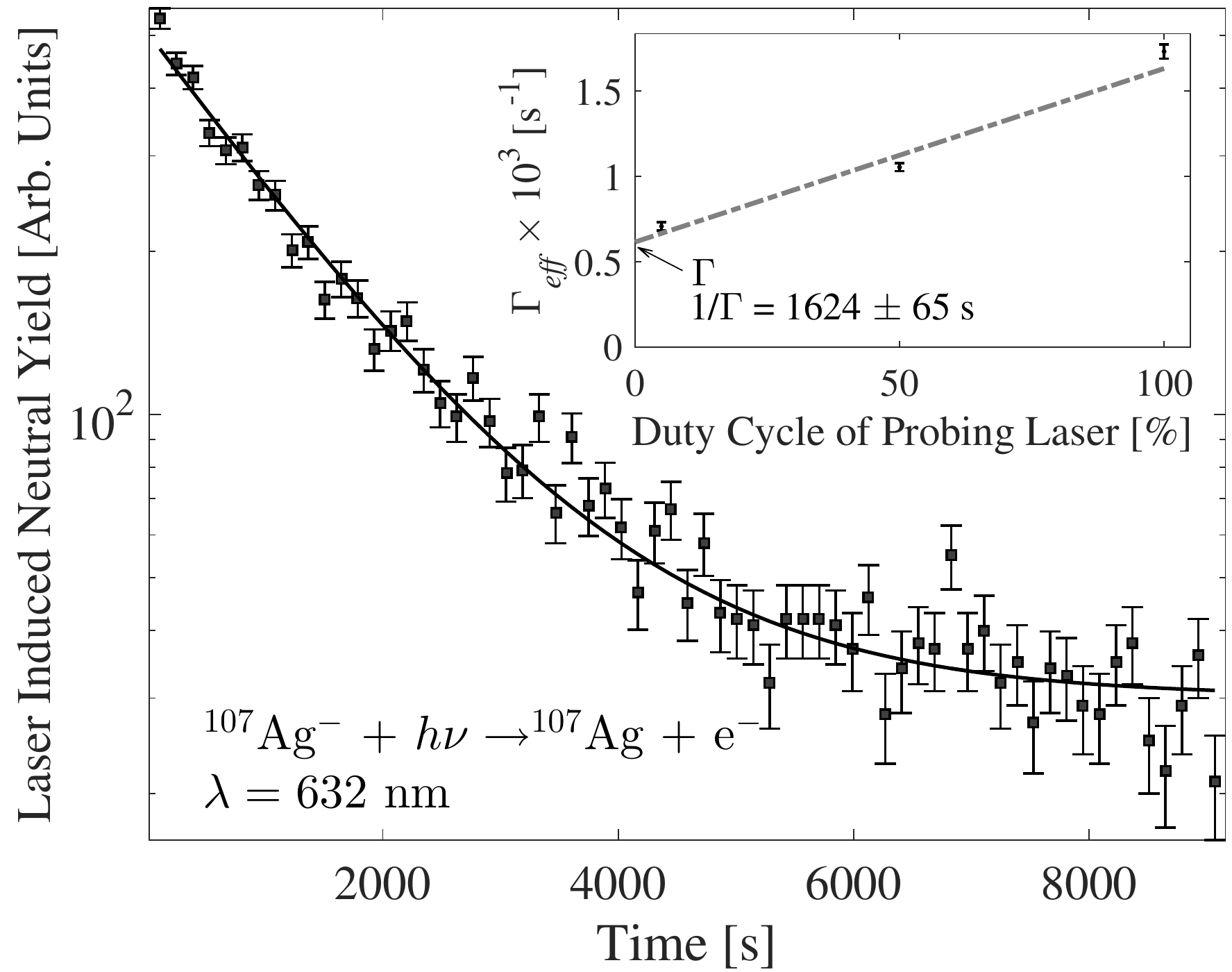}
  
\caption{Recorded neutral yield from photodetachment of silver monomer anions, \ani{Ag}, at $\lambda=$ \SI{632}{\nm} as a function of time with a 5\% laser duty cycle. The solid line is a fit to the data with the sum of a single exponential function and a constant due to the detector dark counts. The inset shows the measured effective decay rate, $\Gamma_\textit{eff}$, as a function of laser duty cycle. The lifetime of the \ani{Ag} beam is inversely proportional to the decay rate, $\Gamma$,  obtained by extrapolating the measured $\Gamma_\textit{eff}$ to zero duty cycle.}
 
 \label{fig:lifetime}
 \end{figure}

A mechanical shutter was used to chop a continuous wave  \SI{632}{\nm} (1.99 eV) laser beam into pulses for the \ani{Ag} storage measurement. This pulse train produced neutral Ag atoms via photodetachment proportional to the number of  \ani{Ag} ions in the ring as a function of time after injection. The laser duty cycle was varied for successive measurements. The laser and ion beams travelled in opposing directions in a collinear configuration in the straight section of the ring. In this configuration, the laser first passed through the glass plate in front of the detector and then interacted with the \ani{Ag} beam. The neutral Ag atoms from the photodetachment process were registered by the MCP. The neutral yield recorded with a laser duty cycle of 5\% is shown in Fig.~\ref{fig:lifetime}. 

The measured effective decay rate $\Gamma_\textit{eff}$ contains contributions from laser-induced photodetachment losses. Therefore the effective decay rate, $\Gamma_\textit{eff}$, was measured as a function of laser duty cycle as shown in the inset in Fig.~\ref{fig:lifetime}. By extrapolating to zero laser duty cycle, a decay rate, $\Gamma$, gave a storage lifetime of $1/\Gamma$ = \SI{1624\pm65}{\s} for \SI{10}{keV} \ani{Ag} ions. Furthermore, the ion beam storage time is expected to increase with cluster size for fixed storage energy, as applied here, and thus ion beam loss due to residual gas collisions during the first 60 seconds (the maximum cluster decay measuring time) of storage is negligibly small.

\begin{figure}
\centering
\includegraphics[width=1\columnwidth]{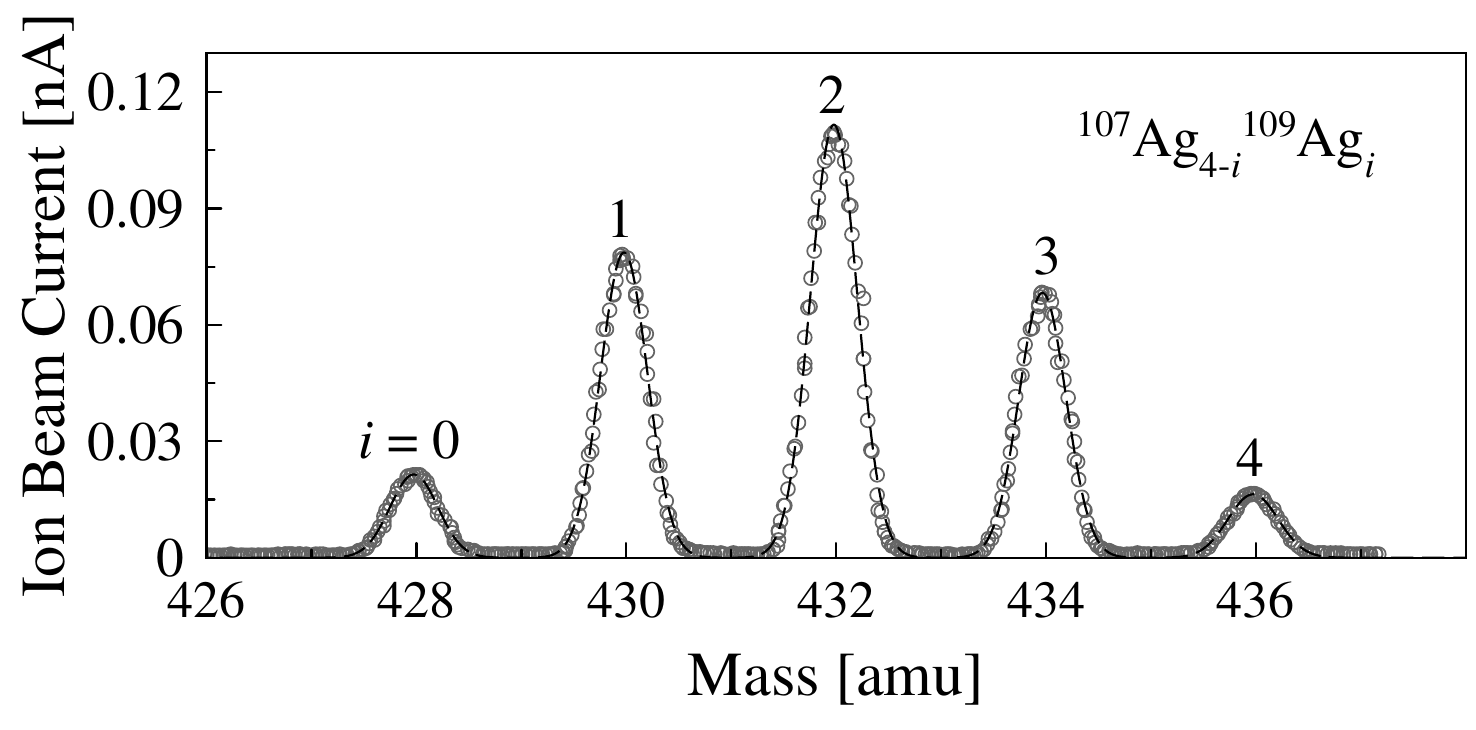}
	
\caption{Recorded mass spectra for the \clu{Ag}{4} anions (open circles).  The peak label $i$ indicates the number of $^{109}$Ag atoms in the clusters. The abundances of the masses were determined from gaussian fits to the peaks (dashed lines) and were found to agree with those expected from the natural abundances within a per cent.}	

\label{fig:mass}
\end{figure}

\subsection {Are there hydrogenated silver cluster anions in the \clu{Ag}{4} beams?}
Silver has two naturally occurring isotopes of mass 107 and 109 amu with close to equal abundances, so there are $n+1$ possible isotopologues for \clu{Ag}{n}. A  mass spectrum for \clu{Ag}{4}, generated by scanning the analyzing magnet and measuring the ion current on a Faraday cup located after the magnet, is shown in Fig. \ref{fig:mass}. The relations between the ion beam currents at masses 428, 430, 432, 434, and 436\,amu agree to within a per cent of those expected from the natural abundances. We found the measured time dependences of the neutral yields due to stored \clu{Ag}{4} beams with different masses (428, 430, and 436\,amu) to be the same within the experimental uncertainty.

Contamination of, for example, the \clu{Ag}{4} beam by \cluH{Ag}{4} ions, could potentially contribute to the measured neutral yields. However, Fig. \ref{fig:mass} shows well separated peaks for the different masses of the \clu{Ag}{4} clusters, suggesting there is no or very little contamination of the ion beams with \cluH{Ag}{4}. This was further investigated by means of a measurement of the \clu{Ag}{4} cluster where the mass of the ions was chosen (by a slight shift of the setting of the analyzing magnet) to be on the high mass shoulder of the main peak, where we expect larger relative contributions to the ion beam from \cluH{Ag}{4}, if at all present. By comparing to a measurement of the time dependence of the neutral yield where the analyzing magnet was set to the center of the \clu{Ag}{4} mass peak, we concluded that \cluH{Ag}{n} contaminations were sufficiently small to be neglected.

\subsection{Measurements with \clu{Ag}{4} -- \clu{Ag}{7}: Time windows and cluster currents}
The neutral yield from \clu{Ag}{4} and \clu{Ag}{5} clusters were measured using several time windows of up to 60\,s. The emission of neutral particles from \clu{Ag}{6} and \clu{Ag}{7} were measured for only a single \SI{100}{ms} time window. A low ion current of a few pA was used for the shortest measurement time window of 100 ms; using a low current prevents saturation of the detector and effectively eliminates the effects of beam losses due to ion-ion interactions~\cite{Schmidt2013}. Higher currents of a few nA were used for longer time windows in cases where the signal rates per molecule were much lower. The different data sets were then combined using an overlapping time region where the data sets are unaffected by saturation. In the results presented in Figs.~\ref{fig:Ag4andAg5} and \ref{fig:FigAg67} we have added data from many injections. The first few points in each figure are the accumulated counts from individual turns of the ion bunch in the ring. Data for all later times are sorted into time bins increasing linearly in width with time so that these data points are equidistant on a logarithmic scale.

\section{Modelling the decay of excited cluster anions}\label {Decay Models}

An excited silver cluster anion, (\clu{Ag}{n})$^{*}$, can decay via the following channels,
\[
  (\text{Ag}_{n}^{-})^{*} \rightarrow \begin{cases}
               \text{Ag}_{n} + e^{-}  \qquad \qquad \text{electron detachment}  \\
               \text{Ag}_{n-m}^{-} + \text{Ag}_{m} \qquad  \text{fragmentation}^\dagger \\
               \text{Ag}_{n}^{-} + hv \qquad \qquad \text{radiative cooling}\\
            \end{cases}          
\]
\hspace{4.7 cm} $^\dagger m=1,...,n-1$\\

The electron detachment and fragmentation processes lead to the production of neutrals and contribute to the measured signal. Electron detachment processes, in the time range relevant here ($\mu$s or longer), will proceed via vibrational autodetachment (VAD) where vibrational energy is transferred to the electron. Radiative cooling processes lower the internal energy of the ion, which remains stored in the ring, with reduced probabilities for fragmentation or electron detachment. The time-dependent total neutralization rate, $R(t)$ following the injection of a bunch of $N$ non-interacting ions with an initial distribution $g(E)$ of internal excitation energies $E$ is

\begin{equation}
R(t) = N \int_{0}^{\infty} g(E)k_\text{neutral}(E)e^{-k_\text{tot}(E)t} \text{d}E.
\label{eq:Rate}
\end{equation}

Here, the neutral particle production rate constant, $k_\text{neutral}(E)=k_\text{VAD}(E)+k_\text{frag}(E)$, and the total decay rate constant, $k_\text{tot}(E)=k_\text{neutral}(E)+k_\text{rad}(E)$ are expressed in terms of the rates for vibrational autodetachment ($k_\text{VAD}$), fragmentation ($k_\text{frag}$), and radiative cooling ($k_\text{rad}$). In Eq.~\ref{eq:Rate}, we assume that a single photon emission event effectively gives $k_\text{neutral}=0$\,$\text{s}^{-1}$ for all later times. The radiative cooling rate, $k_\text{rad}$, is expected to vary much more slowly with $E$ than $k_\text{neutral}(E)$ and may then be treated as a constant. 

As mentioned above, the time dependence of $R(t)$ is in many cases simple and can be characterized through

\begin{equation}
R(t)  \propto t^{-1+\delta}
\label{eq:powerlaw}
\end{equation}

\noindent where $\delta$ can be positive or negative and is often small~\cite{Froese2011, Fedor2005, Andersen2003_b}. 

A strict $t^{-1}$ behavior follows from Eq.~\ref{eq:Rate} provided $k_\text{tot}$ is a sufficiently rapidly increasing function of $E$ and provided that $g(E)$ does not change significantly with $E$ over the (narrow) range of internal energies involved. Then, for $k_\text{rad}(E)<< k_\text{tot}(E)$ the function $k_\text{tot}(E)t e^{-k_\text{tot}(E)t}$ ($\approx$ $k_\text{neutral}(E)t e^{-k_\text{tot}(E)t}$) is strongly peaked at its maximum value at $E=E_{max}$ and $R(t)$ is proportional to $g(E_{max})/t$ (to see this think of $k_\text{neutral}(E)t e^{-k_\text{tot}(E)t}$ as a delta function at $E=E_{max}$ and do the integral in Eq.~\ref{eq:Rate}). Since $k_\text{tot}(E)$ increases strongly with increasing $E$, the initial distribution $g(E)$ is depleted from the high energy side. That is, ions with higher internal energy decay first. As a consequence $E_{max}$ decreases with time but when $g(E)$ is constant, the value of $g(E_{max})$ does not change with time yielding the $t^{-1}$ power law~\cite{Hansen2001}. For large systems, such as amino acids, the absolute value of $\delta$ has typically been found to be smaller than 0.1~\cite{Andersen2003_b}. For small systems $|\delta|$ values of up to 1 have been reported~\cite{Froese2011, Hansen2017, Breitenfeldt2016, Menk2014}.

Radiative cooling will lower the internal energies of the stored ions. These photon emission processes will not produce neutrals but may quench the power law decay at times $t$ approaching a characteristic time $\tau$, when $k_\text{rad}(E)$ is no longer insignificant compared to $k_\text{tot}(E)$.  The neutralization rate including radiative cooling processes, provided that $k_{neutral}$ is effectively negligible after the emission of the photon, is given by~\cite{Hansen2017}

\begin{equation}
 R(t) \propto t^{-1+\delta} e^{-t/\tau},
\label{eq:rate}
\end{equation}

\noindent where $\tau$ is the characteristic photon emission time. 

To enable quantitative comparisons with our experimental data we calculate the rate constants for electron detachment and fragmentation processes based on detailed balance considerations. The calculated rate constants and a constant value of $g(E)$ will then be used in Eq.~\ref{eq:Rate} to calculate the total rates for neutral particle production. 

The rate constant for electron detachment is taken to be~\cite{Weisskopf1937, Hansen2013}

\begin{equation}
k_\text{VAD}\left( E,\varepsilon\right) = \dfrac{2m_e}{\pi^{2}\hbar^{3}} \varepsilon \sigma_{L}(\varepsilon)  \dfrac{\rho^{(0)}\left(E-E_{a}-\varepsilon \right) }{\rho^{(-)}(E)},
\label{eq:k_VAD}
\end{equation}

\noindent where $E$ is the internal excitation energy, $\varepsilon$ the kinetic energy of the emitted electron, $\sigma_{L}(\varepsilon)$ is the Langevin cross section for electron attachment to the neutral silver cluster, $m_{e}$ is the mass of the electron, and $E_a$ is the electron affinity of the corresponding neutral cluster. $\rho^{(0)}$ and $\rho^{(-)}$ are the vibrational level densities of the Ag$_{n}$ product (with internal energy, $E-E_a-\varepsilon$) and of the initial \clu{Ag}{n} state (with internal energy $E$), respectively. The factor of 2 is due to the spin degeneracy of the emitted electron. The level densities can be determined using the Beyer-Swinehart algorithm~\cite{Beyer1973} where vibrational frequencies, electron affinities and dissociation thresholds are required as input.

In analogy with Eq.~\ref{eq:k_VAD} we express the rate constant, $k_\text{frag}(E, \varepsilon)$, for a fragmentation process in which the cluster parent anion (\clu{Ag}{n}) emits a neutral silver monomer or dimer as
\begin{equation}
k_\text{frag}\left( E,\varepsilon\right) = \dfrac{\gamma\mu}{\pi^{2}\hbar^{3}} \varepsilon\sigma_{c}  \dfrac{\rho^{(d)}\left(E-E_{D}-\varepsilon \right) }{\rho^{(p)}(E)}.
\label{eq:k_Frag}
\end{equation}

\noindent Here, $E$ is again the excitation energy of the parent ion ($p$), $\varepsilon$ is the sum of the kinetic energy release and any internal excitation energy of the emitted particle, $E_{D}$ is the dissociation energy for the given fragmentation channel and $\mu$ the reduced mass of the two fragments, and $\sigma_c$ is the cross section for the reverse process where the parent system is formed from the fragments. The level densities of the parent ($p$) and daughter ($d$) systems are $\rho^{(p)}$ and $\rho^{(d)}$ with internal energies $E$ and $E-E_{D}-\varepsilon$, respectively. The degeneracy of the emitted fragment is denoted by $\gamma$.

The summed electron detachment and fragmentation rates are

\begin{equation}
\begin{aligned}
k_\text{VAD}\left(E\right) =& \int k_\text{VAD}\left (E, \varepsilon \right) \text{d}\varepsilon \qquad \text{ and} \\
k_\text{frag}\left(E\right) =& \int k_\text{frag}\left (E, \varepsilon \right) \text{d}\varepsilon \text{,}
\label{eq:k_sums}
\end{aligned}
\end{equation}

\noindent respectively.

\section{Results and Discussion}

\subsection{\clu{Ag}{4} and \clu{Ag}{5}: Experimental data}

\begin{figure}
\centering	
\includegraphics[width=\columnwidth]{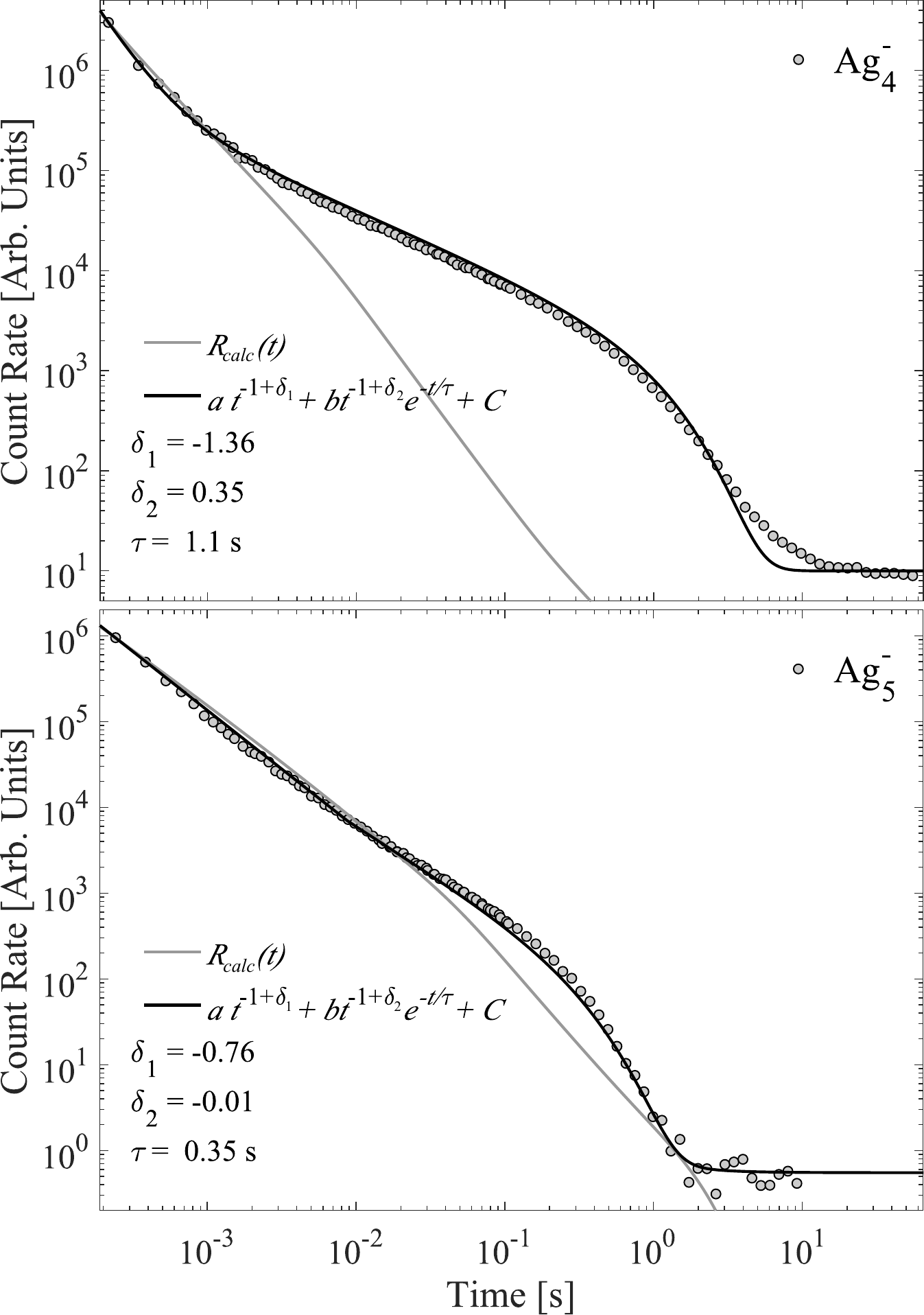}	
\caption{Neutral particle yield from stored beams of \clu{Ag}{4}  and \clu{Ag}{5} as a function of time after the ions are formed in the source. The solid black lines are curves described by Eq.~\ref{eq:rate2} with parameter values as indicated. The grey lines are the calculated neutral emission rates $R_{calc}(t)$ with no radiative cooling included ($k_\text{rad} = 0 \text{\,s}^{-1}$) and where rotational excitations are neglected.}
\label{fig:Ag4andAg5}
\end{figure}

The measured neutral particle yields from the stored \clu{Ag}{4} and \clu{Ag}{5} anion beams are shown as functions of time in log-log plots in Fig.~\ref{fig:Ag4andAg5}. At short times, steep linear behavior appears in both plots corresponding to power laws $t^{-1+\delta}$ with large values of $|\delta|$. In both cases (\clu{Ag}{4} and \clu{Ag}{5}), this changes to a less steep slope over time ranges of a few tens or hundreds of milliseconds. Finally, the curves bend down until reaching levels that slowly decrease. The latter are slow decays of the stored ion beams due to residual-gas collisions (see Section~\ref{beam storage}). Similar behavior (i.e. different exponents in different time ranges) have been reported for the spontaneous decay from small copper cluster anions of size $n =$~3~--~6~\cite{Hansen2017,Breitenfeldt2016}. 

A fitted function of the form

\begin{equation}
R(t) = a t^{-1+\delta_1} + b t^{-1+\delta_2} e^{-t/\tau} + C
\label{eq:rate2}
\end{equation}

\noindent is shown together with the experimental data in Fig.~\ref{fig:Ag4andAg5}. The first term in Eq.~\ref{eq:rate2} is a power law and the second term an exponentially quenched power law (as in Eq.~\ref{eq:rate}). The constant $C$ is the contribution from residual gas collisions - the decay of the beam is very slow (see Section~\ref{beam storage}). The parameters $\delta_1$, $\delta_2$ and $\tau$ used to fit Eq.~\ref{eq:rate2} to the experimental data are listed in Fig.~\ref{fig:Ag4andAg5}. Two-component power laws may indicate that two different classes of \clu{Ag}{n} ions are simultaneously stored in the ring.

In previous studies of small silver clusters ~\cite{Hansen2001} an exponential cut-off with a characteristic time of $\tau=$~\SI{5}{ms} was reported for \clu{Ag}{4} and was then ascribed to radiative cooling~\cite{Hansen2001}. We do not see an exponential cut-off in this time range in the present data for \clu{Ag}{4}. We do, however, observe a characteristic exponential cut-off time of $\tau=$~\SI{1.1}{s}. Here we note that conditions for the observation of the signal at long times are more favourable in DESIREE than in earlier experiments. In the previous silver cluster studies~\cite{Hansen2001}, no deviation from a power law behavior was observed for storage times up to \SI{50}{ms} for \clu{Ag}{5}. This is consistent with the measurements presented here, where an exponential cut-off with a characteristic time of $\tau=$~\SI{0.35}{s} is deduced. The present characteristic times are of the same order of magnitude as those found for small copper clusters stored in DESIREE~\cite{Hansen2017} ($\tau=$~\SI{0.83}{s} for  \clu{Cu}{4} and $\tau=$~\SI{0.18}{s} for \clu{Cu}{5}).

\begin{figure}
\centering
\includegraphics[width=\columnwidth]{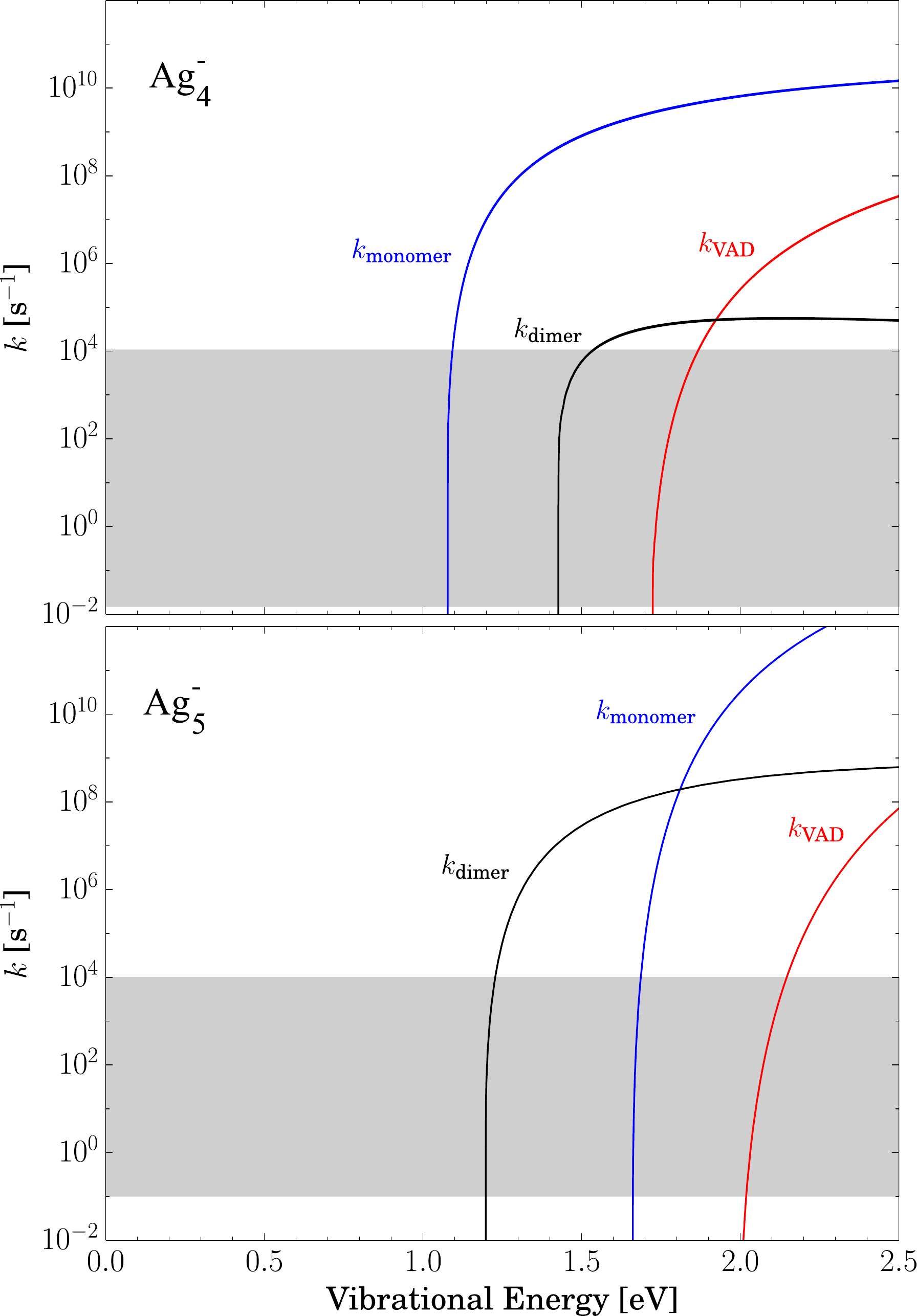}	
\caption{The calculated rate constants for \clu{Ag}{4} and \clu{Ag}{5} for vibrational autodetachment (VAD) and fragmentation resulting in the loss of a neutral monomer or neutral dimer as functions of the internal vibrational energies of the parent anions. In these calculations rotational excitations are neglected. The grey shaded area shows the rate constants that correspond to the experimental time window.}
\label{fig:Ag4andAg5_k_rates}
\end{figure}

\subsection{\clu{Ag}{4} and \clu{Ag}{5}: Rate calculations and comparisons}

We calculate the rates for electron detachment and for fragmentation through emission of neutral monomers and neutral dimers separately for both cluster sizes following the procedures outlined in Section~\ref{Decay Models}. Rotational excitations are neglected in these calculations.

Several parameters are required in each case and for the detachment processes we first calculate $\sigma_{L}(\varepsilon)$, the Langevin electron capture cross section for both cluster sizes using a polarizability of \SI{15.8}{\angstrom^3}. This value was taken from reference~\cite{Breitenfeldt2016} where it was extrapolated from experimental data for \clu{Cu}{4}. The Langevin cross section assumes unit sticking probability and the calculated rate constant for vibrational autodetachment (see Eq.~\ref{eq:k_VAD})  will thus serve as an upper limit within the present model. For the fragmentation processes a geometric value of  $\sigma_c=$\SI{1}{\angstrom^2} was used for the formation cross section (see Eq.~\ref{eq:k_Frag}).

The harmonic vibrational frequencies, electron affinities and dissociation thresholds needed for the level densities were calculated by means of density functional theory (DFT) at the B3LYP/LANL2DZ level of theory. The resulting electron affinities for Ag$_{4}$ and Ag$_{5}$ were then found to be \SI{1.72}{eV} and \SI{2.01}{eV} respectively. The dissociation thresholds for monomer and dimer loss were calculated to be \SI{1.08}{eV} and \SI{1.43}{eV} respectively for \clu{Ag}{4} and \SI{1.66}{eV} and \SI{1.20}{eV} respectively for \clu{Ag}{5}. A degeneracy of $\gamma =2$ was used when calculating the fragmentation rates from Eq.~\ref{eq:k_Frag}.  Atomic silver has a spin-1/2 ground state, so the degeneracy value for monomer emission is 2. For dimer emission the spin degeneracy could be 1 or 3 depending on whether the dimer is in a singlet or triplet state. Here we use a rough average value of 2. The radiative decay rate, $k_\text{rad}$ was set to zero.

In Fig.~\ref{fig:Ag4andAg5_k_rates}, we show calculated rates for electron detachment via VAD and fragmentation of \clu{Ag}{4} and \clu{Ag}{5}. The grey shaded areas in the plots indicate rates that correspond to the detection time window of the present ion beam storage experiment. The total neutralization rates for \clu{Ag}{4} and \clu{Ag}{5} resulting from the calculated data in Fig.~\ref{fig:Ag4andAg5_k_rates} are shown as grey lines in Fig.~\ref{fig:Ag4andAg5}.

The rate constants calculated for \clu{Ag}{4} (upper panel Fig.~\ref{fig:Ag4andAg5_k_rates}) indicate that monomer loss (i.e. fragmentation) is the dominant process at all times where we have a decay signal in the experiment and that all neutrals detected in the measurement are silver atoms from this dissociation process. For \clu{Ag}{5} the calculations (lower panel Fig.~\ref{fig:Ag4andAg5_k_rates}) indicate that fragmentation through dimer loss is the only process contributing to the measured yield of neutrals. The corresponding calculated neutral rates (using the method described in Section~\ref{Decay Models} and shown in Fig.~\ref{fig:Ag4andAg5}) reproduce the surprisingly steep slopes at early times in both cases (\clu{Ag}{4} and \clu{Ag}{5}), but fall far below the measured rate at later times.  As the initial strong deviations from $t^{-1}$  are reproduced in the calculations, which assume constant values for the initial internal energy distributions $g(E)$, specific initial conditions are likely {\it not} the cause of these deviations. Furthermore, as was argued in Section~\ref{Decay Models} and in several earlier studies of the power-law decay phenomenon~\cite{Hansen2001}, $t^{-1}$ follows from simple arguments when $g(E)$ is assumed constant and when $k_\text{tot}(E)t e^{-k_\text{tot}(E)t}$ is assumed to be a delta function. This indicates that the large $\delta$ values for early times may be connected to the explicit energy dependence of the decay rate.

\subsection{\clu{Ag}{4} and \clu{Ag}{5}: High-$J$ contributions?}

While the model works relatively well for short times it appears that ensembles of more slowly decaying clusters are present in the stored \clu{Ag}{4} and \clu{Ag}{5} beams and that these could dominate the neutralisation signals on long time scales. This other class could be molecules in high rotational states as discussed in a recent study of \clu{Cu}{n}~\cite{Hansen2017}. There, electron detachment was the lowest energy decay channel for low $J$, while fragmentation was energetically more favourable at higher $J$. It was then speculated that this could be related to the two-component shape of the decay curve for some of the small copper clusters~\cite{Hansen2017}. 

\begin{figure}
	\includegraphics[width=\columnwidth]{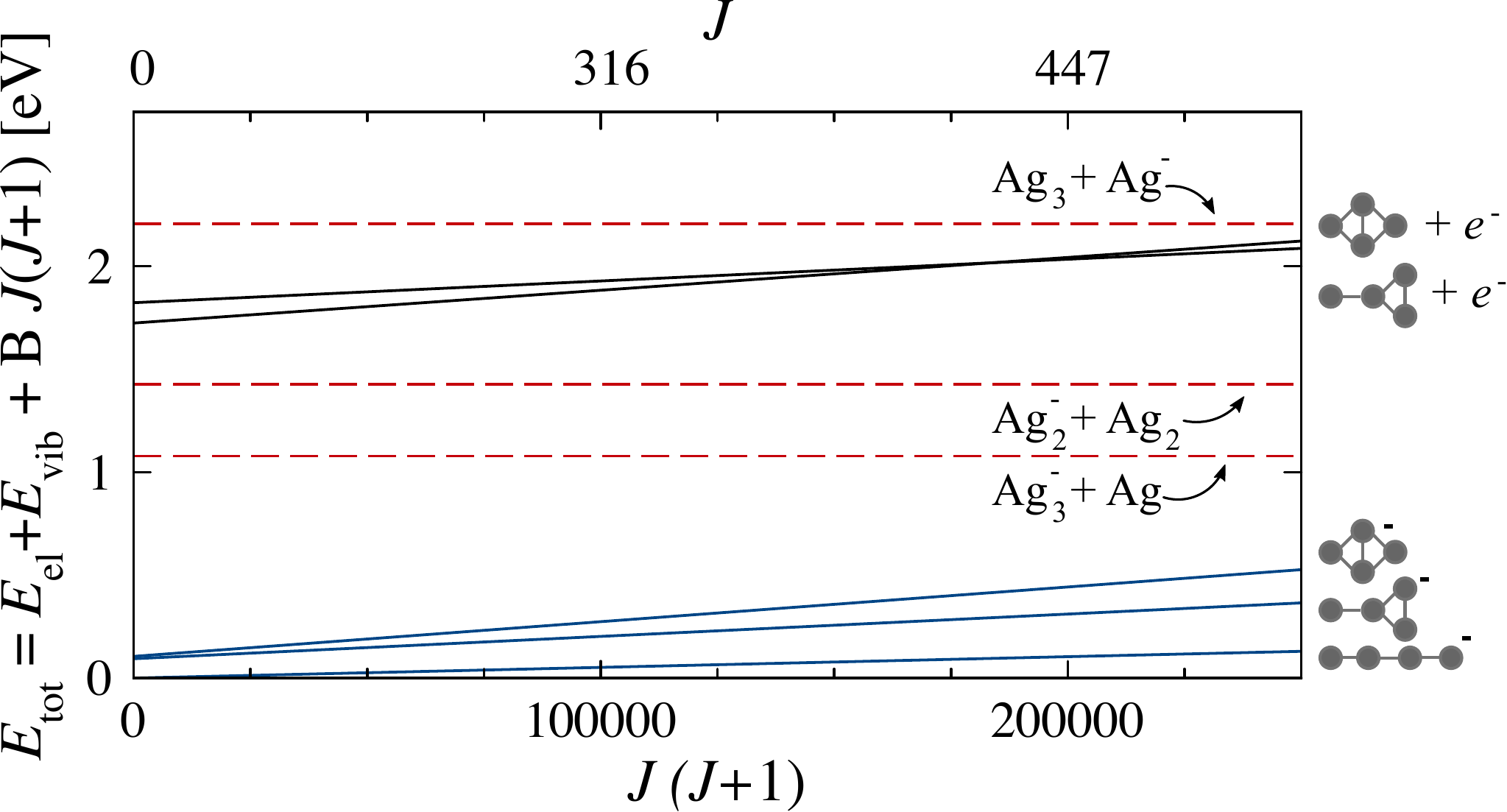}
	\caption{Yrast plot for \clu{Ag}{4}. Shown are the calculated relative excitation energies (blue lines) and corresponding electron detachment limits (black lines) as functions of $J(J+1)$, where $J$ is the rotational angular momentum quantum number. The lowest energy fragmentation channels (dashed red lines) are calculated for $J=0$. In principle these dashed lines should have small non zero slopes as the rotational barrier height increases with $J$. Full dynamics simulations would be required to properly describe this for all geometries, which is beyond the scope of this work. Rotations along the axis with the smallest value of B were assumed to give lower limits to the excitation energies. The excited vibrational states, which will extend from the lowest state for each structure, are not indicated.}
\label{fig:Ag4_yrast}
\end{figure}

For \clu{Ag}{4} and \clu{Ag}{5}, the threshold energy for fragmentation is lower than the threshold energy for electron detachment for much wider ranges of angular momenta. For \clu{Ag}{4}, this is illustrated in Fig.~\ref{fig:Ag4_yrast}, where the minimum excitation energies needed for, fragmentation and electron detachment are shown as functions of $J(J+1)$. These energies were calculated by means of density functional theory (DFT) for stable structural isomers. The \clu{Ag}{4} isomers have rhombic, Y-shaped and linear forms and the energies of their vibrational ground states as a function of $J(J+1)$ are shown as solid blue lines in Fig.~\ref{fig:Ag4_yrast}. The relevant electron detachment limits (solid black lines) lie above the two lowest fragmentation limits (red dashed lines) over the whole range of $J$ values covered in Fig~\ref{fig:Ag4_yrast}. Unlike the situation for the copper clusters,  the dominating decay mechanism is the same irrespective of the value of the total angular momentum quantum number within certain bounds. This, however, does not exclude the possibility that ions with very different values of $J$ can decay through fragmentation on very different time scales and that the slow components that are not described by the present model could be due to clusters with high values of $J$.

\subsection{\clu{Ag}{6} and \clu{Ag}{7}}

\begin{figure}
	\includegraphics[width=\columnwidth]{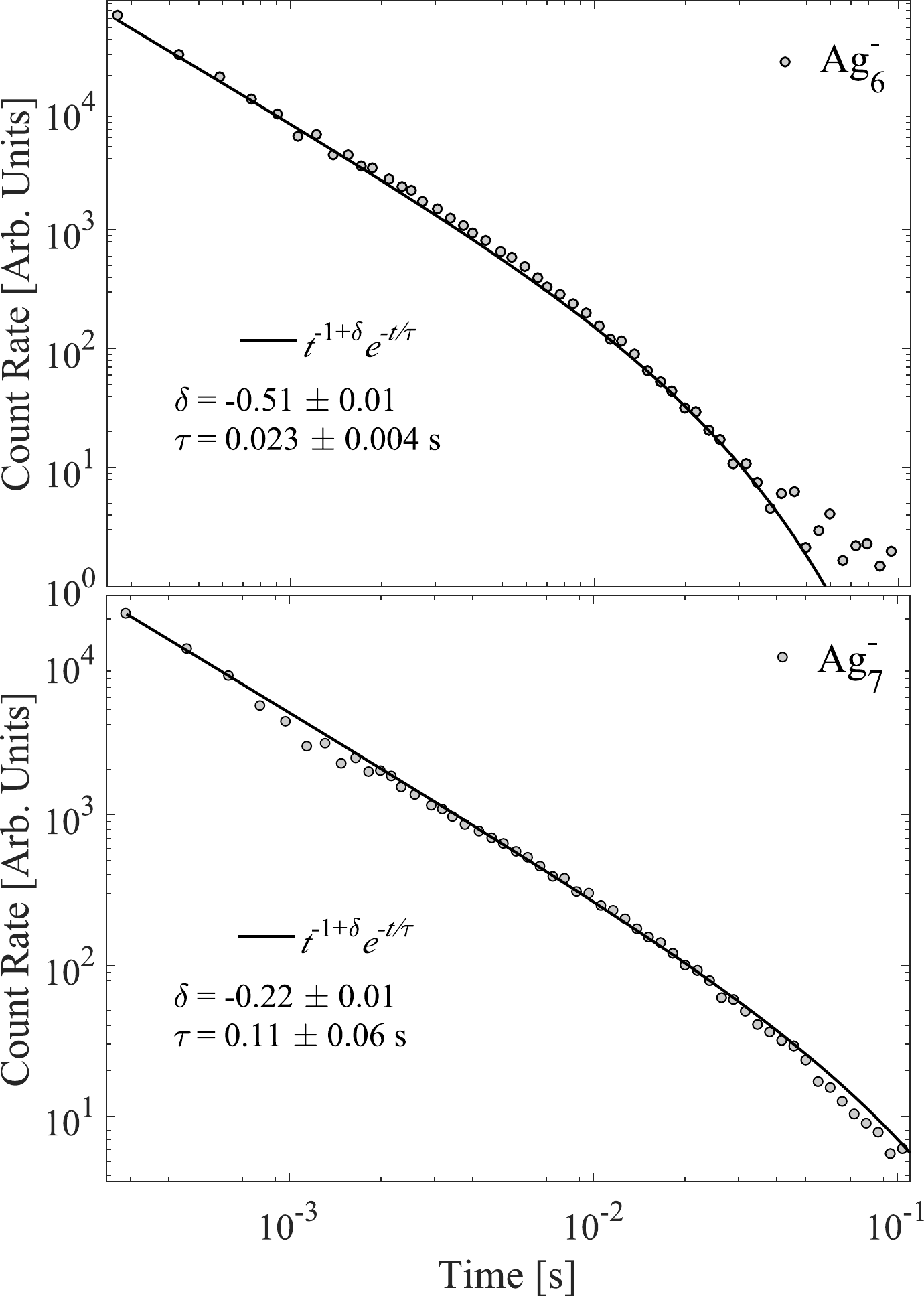}
		
	\caption{Neutral particle yield from stored beams of silver clusters \clu{Ag}{6} and \clu{Ag}{7} as a~function of time.}
	\label{fig:FigAg67}
\end{figure}

The neutral particle yields from \clu{Ag}{6} and \clu{Ag}{7} ions are shown as functions of time in Fig.~\ref{fig:FigAg67}. The decay curves for both these cluster ions show similar behavior, a power law at early times followed by an exponential drop-off at later times. This shape is indicative of an ensemble of cluster anions with a broad internal energy distribution where the decay is dominated by electron detachment and/or fragmentation at early times, whilst at later times the neutral particle emission yield is quenched by e.g.~radiative cooling processes. Eq.~\ref{eq:rate} describes this behavior and was used to fit the experimental data to obtain the exponential cut off times of $\tau=23 \pm 4$ and $\tau=110 \pm 60$\,ms for the \clu{Ag}{6} and \clu{Ag}{7} anions respectively. The latter value agrees well with the previously reported value of $ \sim $\SI{100}{ms} for \clu{Ag}{7}, while the former value is dramatically longer than the $\sim $\SI{5}{ms} reported for \clu{Ag}{6} in the previous study \cite{Hansen2001}. For these more complex systems (with many possible isomers and dissociation channels) we have not attempted to make any model calculations.

\section{Summary and conclusions} 

In this work we have presented measurements of the decays of small internally hot \clu{Ag}{n}~\ $n=4-7$ anions injected and stored in one of the cryogenic ion beam storage rings of DESIREE. The neutral yields of particles from stored \clu{Ag}{4} and \clu{Ag}{5} anion beams follow $t^{-1+\delta}$ behavior with large $|\delta|$ values at early times but follow less steep power laws at later times. These experimental observations prompted calculations of the rate constants for vibrational detachment and fragmentation of rotational ground state \clu{Ag}{4} and \clu{Ag}{5}. These detailed-balance calculations strongly suggest that fragmentation dominates the decays of \clu{Ag}{4} and \clu{Ag}{5} through the emission of neutral monomers and dimers, respectively. The calculated neutral yields reproduce the measured steep slopes. Since the calculations use a constant value for the initial energy distribution we ascribe the large $|\delta|$ values to inherent properties of the specific clusters rather than to the ion source conditions. Whilst the calculated neutral rates reproduce the early times of the experimental yields they deviate at later times indicating that there are ensembles of more slowly decaying clusters in the \clu{Ag}{4} and \clu{Ag}{5} beams. The origin of the two-component structures in the \clu{Ag}{4} and \clu{Ag}{5} data are not yet clear but could possibly be connected to ions with low and high angular momenta. The \clu{Ag}{6} and \clu{Ag}{7} decay curves are found to follow single power laws with exponential cut offs of $23 \pm 6$~ms and $110 \pm 60$~ms respectively.

In future experimental campaigns we plan to perform coincidence measurements between neutral and charged fragments in order to conclusively determine if fragmentation or electron detachment dominate in the decay of small metal clusters.

\section*{Acknowledgements}
This work was performed at the Swedish National Infrastructure, DESIREE (Swedish Research Council contract No. 2017-00621). It was further supported by the Swedish Research Council (grant numbers 2014-4501, 2015-04990, 2016-04181, 2016-06625). M. K. acknowledges financial support from the Mobility Plus Program (Project No. 1302/MOB/IV/2015/0) funded by the Polish Ministry of Science and Higher Education.

\bibliography{../../../../../Ag_paper}

\begin{thebibliography}{36}%
\makeatletter
\providecommand \@ifxundefined [1]{%
 \@ifx{#1\undefined}
}%
\providecommand \@ifnum [1]{%
 \ifnum #1\expandafter \@firstoftwo
 \else \expandafter \@secondoftwo
 \fi
}%
\providecommand \@ifx [1]{%
 \ifx #1\expandafter \@firstoftwo
 \else \expandafter \@secondoftwo
 \fi
}%
\providecommand \natexlab [1]{#1}%
\providecommand \enquote  [1]{``#1''}%
\providecommand \bibnamefont  [1]{#1}%
\providecommand \bibfnamefont [1]{#1}%
\providecommand \citenamefont [1]{#1}%
\providecommand \href@noop [0]{\@secondoftwo}%
\providecommand \href [0]{\begingroup \@sanitize@url \@href}%
\providecommand \@href[1]{\@@startlink{#1}\@@href}%
\providecommand \@@href[1]{\endgroup#1\@@endlink}%
\providecommand \@sanitize@url [0]{\catcode `\\12\catcode `\$12\catcode
  `\&12\catcode `\#12\catcode `\^12\catcode `\_12\catcode `\%12\relax}%
\providecommand \@@startlink[1]{}%
\providecommand \@@endlink[0]{}%
\providecommand \url  [0]{\begingroup\@sanitize@url \@url }%
\providecommand \@url [1]{\endgroup\@href {#1}{\urlprefix }}%
\providecommand \urlprefix  [0]{URL }%
\providecommand \Eprint [0]{\href }%
\providecommand \doibase [0]{http://dx.doi.org/}%
\providecommand \selectlanguage [0]{\@gobble}%
\providecommand \bibinfo  [0]{\@secondoftwo}%
\providecommand \bibfield  [0]{\@secondoftwo}%
\providecommand \translation [1]{[#1]}%
\providecommand \BibitemOpen [0]{}%
\providecommand \bibitemStop [0]{}%
\providecommand \bibitemNoStop [0]{.\EOS\space}%
\providecommand \EOS [0]{\spacefactor3000\relax}%
\providecommand \BibitemShut  [1]{\csname bibitem#1\endcsname}%
\let\auto@bib@innerbib\@empty
\bibitem [{\citenamefont {Thomas}\ \emph {et~al.}(2011)\citenamefont {Thomas},
  \citenamefont {Schmidt}, \citenamefont {Andler}, \citenamefont
  {Bj{\"o}rkhage}, \citenamefont {Blom}, \citenamefont {Br{\"a}nnholm},
  \citenamefont {B{\"a}ckstr{\"o}m}, \citenamefont {Danared}, \citenamefont
  {Das}, \citenamefont {Haag}, \citenamefont {Halld{\'e}n}, \citenamefont
  {Hellberg}, \citenamefont {Holm}, \citenamefont {Johansson}, \citenamefont
  {K{\"a}llberg}, \citenamefont {K{\"a}llersj{\"o}}, \citenamefont {Larsson},
  \citenamefont {Leontein}, \citenamefont {Liljeby}, \citenamefont
  {L{\"o}fgren}, \citenamefont {Malm}, \citenamefont {Mannervik}, \citenamefont
  {Masuda}, \citenamefont {Misra}, \citenamefont {Orb{\'a}n}, \citenamefont
  {Pa{\'a}l}, \citenamefont {Reinhed}, \citenamefont {Rensfelt}, \citenamefont
  {Ros{\'e}n}, \citenamefont {Schmidt}, \citenamefont {Seitz}, \citenamefont
  {Simonsson}, \citenamefont {Weimer}, \citenamefont {Zettergren},\ and\
  \citenamefont {Cederquist}}]{Thomas2011}%
  \BibitemOpen
  \bibfield  {author} {\bibinfo {author} {\bibfnamefont {R.~D.}\ \bibnamefont
  {Thomas}}, \bibinfo {author} {\bibfnamefont {H.~T.}\ \bibnamefont {Schmidt}},
  \bibinfo {author} {\bibfnamefont {G.}~\bibnamefont {Andler}}, \bibinfo
  {author} {\bibfnamefont {M.}~\bibnamefont {Bj{\"o}rkhage}}, \bibinfo {author}
  {\bibfnamefont {M.}~\bibnamefont {Blom}}, \bibinfo {author} {\bibfnamefont
  {L.}~\bibnamefont {Br{\"a}nnholm}}, \bibinfo {author} {\bibfnamefont
  {E.}~\bibnamefont {B{\"a}ckstr{\"o}m}}, \bibinfo {author} {\bibfnamefont
  {H.}~\bibnamefont {Danared}}, \bibinfo {author} {\bibfnamefont
  {S.}~\bibnamefont {Das}}, \bibinfo {author} {\bibfnamefont {N.}~\bibnamefont
  {Haag}}, \bibinfo {author} {\bibfnamefont {P.}~\bibnamefont {Halld{\'e}n}},
  \bibinfo {author} {\bibfnamefont {F.}~\bibnamefont {Hellberg}}, \bibinfo
  {author} {\bibfnamefont {A.~I.~S.}\ \bibnamefont {Holm}}, \bibinfo {author}
  {\bibfnamefont {H.~A.~B.}\ \bibnamefont {Johansson}}, \bibinfo {author}
  {\bibfnamefont {A.}~\bibnamefont {K{\"a}llberg}}, \bibinfo {author}
  {\bibfnamefont {G.}~\bibnamefont {K{\"a}llersj{\"o}}}, \bibinfo {author}
  {\bibfnamefont {M.}~\bibnamefont {Larsson}}, \bibinfo {author} {\bibfnamefont
  {S.}~\bibnamefont {Leontein}}, \bibinfo {author} {\bibfnamefont
  {L.}~\bibnamefont {Liljeby}}, \bibinfo {author} {\bibfnamefont
  {P.}~\bibnamefont {L{\"o}fgren}}, \bibinfo {author} {\bibfnamefont
  {B.}~\bibnamefont {Malm}}, \bibinfo {author} {\bibfnamefont {S.}~\bibnamefont
  {Mannervik}}, \bibinfo {author} {\bibfnamefont {M.}~\bibnamefont {Masuda}},
  \bibinfo {author} {\bibfnamefont {D.}~\bibnamefont {Misra}}, \bibinfo
  {author} {\bibfnamefont {A.}~\bibnamefont {Orb{\'a}n}}, \bibinfo {author}
  {\bibfnamefont {A.}~\bibnamefont {Pa{\'a}l}}, \bibinfo {author}
  {\bibfnamefont {P.}~\bibnamefont {Reinhed}}, \bibinfo {author} {\bibfnamefont
  {K.-G.}\ \bibnamefont {Rensfelt}}, \bibinfo {author} {\bibfnamefont
  {S.}~\bibnamefont {Ros{\'e}n}}, \bibinfo {author} {\bibfnamefont
  {K.}~\bibnamefont {Schmidt}}, \bibinfo {author} {\bibfnamefont
  {F.}~\bibnamefont {Seitz}}, \bibinfo {author} {\bibfnamefont
  {A.}~\bibnamefont {Simonsson}}, \bibinfo {author} {\bibfnamefont
  {J.}~\bibnamefont {Weimer}}, \bibinfo {author} {\bibfnamefont
  {H.}~\bibnamefont {Zettergren}}, \ and\ \bibinfo {author} {\bibfnamefont
  {H.}~\bibnamefont {Cederquist}},\ }\href@noop {} {\bibfield  {journal}
  {\bibinfo  {journal} {Review of Scientific Instruments}\ }\textbf {\bibinfo
  {volume} {82}},\ \bibinfo {pages} {065112} (\bibinfo {year}
  {2011})}\BibitemShut {NoStop}%
\bibitem [{\citenamefont {Schmidt}\ \emph {et~al.}(2013)\citenamefont
  {Schmidt}, \citenamefont {Thomas}, \citenamefont {Gatchell}, \citenamefont
  {Ros\'{e}n}, \citenamefont {Reinhed}, \citenamefont {L\"{o}fgren},
  \citenamefont {Br\"{a}nnholm}, \citenamefont {Blom}, \citenamefont
  {Bj\"{o}rkhage}, \citenamefont {B\"{a}ckstr\"{o}m}, \citenamefont
  {Alexander}, \citenamefont {Leontein}, \citenamefont {Hanstorp},
  \citenamefont {Zettergren}, \citenamefont {Liljeby}, \citenamefont
  {K\"{a}llberg}, \citenamefont {Simonsson}, \citenamefont {Hellberg},
  \citenamefont {Mannervik}, \citenamefont {Larsson}, \citenamefont {Geppert},
  \citenamefont {Rensfelt}, \citenamefont {Danared}, \citenamefont {Pa\'{a}l},
  \citenamefont {Masuda}, \citenamefont {Halld\'{e}n}, \citenamefont {Andler},
  \citenamefont {Stockett}, \citenamefont {Chen}, \citenamefont
  {K\"{a}llersj\"{o}}, \citenamefont {Weimer}, \citenamefont {Hansen},
  \citenamefont {Hartman},\ and\ \citenamefont {Cederquist}}]{Schmidt2013}%
  \BibitemOpen
  \bibfield  {author} {\bibinfo {author} {\bibfnamefont {H.~T.}\ \bibnamefont
  {Schmidt}}, \bibinfo {author} {\bibfnamefont {R.~D.}\ \bibnamefont {Thomas}},
  \bibinfo {author} {\bibfnamefont {M.}~\bibnamefont {Gatchell}}, \bibinfo
  {author} {\bibfnamefont {S.}~\bibnamefont {Ros\'{e}n}}, \bibinfo {author}
  {\bibfnamefont {P.}~\bibnamefont {Reinhed}}, \bibinfo {author} {\bibfnamefont
  {P.}~\bibnamefont {L\"{o}fgren}}, \bibinfo {author} {\bibfnamefont
  {L.}~\bibnamefont {Br\"{a}nnholm}}, \bibinfo {author} {\bibfnamefont
  {M.}~\bibnamefont {Blom}}, \bibinfo {author} {\bibfnamefont {M.}~\bibnamefont
  {Bj\"{o}rkhage}}, \bibinfo {author} {\bibfnamefont {E.}~\bibnamefont
  {B\"{a}ckstr\"{o}m}}, \bibinfo {author} {\bibfnamefont {J.~D.}\ \bibnamefont
  {Alexander}}, \bibinfo {author} {\bibfnamefont {S.}~\bibnamefont {Leontein}},
  \bibinfo {author} {\bibfnamefont {D.}~\bibnamefont {Hanstorp}}, \bibinfo
  {author} {\bibfnamefont {H.}~\bibnamefont {Zettergren}}, \bibinfo {author}
  {\bibfnamefont {L.}~\bibnamefont {Liljeby}}, \bibinfo {author} {\bibfnamefont
  {A.}~\bibnamefont {K\"{a}llberg}}, \bibinfo {author} {\bibfnamefont
  {A.}~\bibnamefont {Simonsson}}, \bibinfo {author} {\bibfnamefont
  {F.}~\bibnamefont {Hellberg}}, \bibinfo {author} {\bibfnamefont
  {S.}~\bibnamefont {Mannervik}}, \bibinfo {author} {\bibfnamefont
  {M.}~\bibnamefont {Larsson}}, \bibinfo {author} {\bibfnamefont {W.~D.}\
  \bibnamefont {Geppert}}, \bibinfo {author} {\bibfnamefont {K.~G.}\
  \bibnamefont {Rensfelt}}, \bibinfo {author} {\bibfnamefont {H.}~\bibnamefont
  {Danared}}, \bibinfo {author} {\bibfnamefont {A.}~\bibnamefont {Pa\'{a}l}},
  \bibinfo {author} {\bibfnamefont {M.}~\bibnamefont {Masuda}}, \bibinfo
  {author} {\bibfnamefont {P.}~\bibnamefont {Halld\'{e}n}}, \bibinfo {author}
  {\bibfnamefont {G.}~\bibnamefont {Andler}}, \bibinfo {author} {\bibfnamefont
  {M.~H.}\ \bibnamefont {Stockett}}, \bibinfo {author} {\bibfnamefont
  {T.}~\bibnamefont {Chen}}, \bibinfo {author} {\bibfnamefont {G.}~\bibnamefont
  {K\"{a}llersj\"{o}}}, \bibinfo {author} {\bibfnamefont {J.}~\bibnamefont
  {Weimer}}, \bibinfo {author} {\bibfnamefont {K.}~\bibnamefont {Hansen}},
  \bibinfo {author} {\bibfnamefont {H.}~\bibnamefont {Hartman}}, \ and\
  \bibinfo {author} {\bibfnamefont {H.}~\bibnamefont {Cederquist}},\
  }\href@noop {} {\bibfield  {journal} {\bibinfo  {journal} {Review of
  Scientific Instruments}\ }\textbf {\bibinfo {volume} {84}},\ \bibinfo {pages}
  {055115} (\bibinfo {year} {2013})}\BibitemShut {NoStop}%
\bibitem [{\citenamefont {Reinhed}\ \emph {et~al.}(2009)\citenamefont
  {Reinhed}, \citenamefont {Orb\'an}, \citenamefont {Werner}, \citenamefont
  {Ros\'en}, \citenamefont {Thomas}, \citenamefont {Kashperka}, \citenamefont
  {Johansson}, \citenamefont {Misra}, \citenamefont {Br\"annholm},
  \citenamefont {Bj\"orkhage}, \citenamefont {Cederquist},\ and\ \citenamefont
  {Schmidt}}]{Reinhed2009}%
  \BibitemOpen
  \bibfield  {author} {\bibinfo {author} {\bibfnamefont {P.}~\bibnamefont
  {Reinhed}}, \bibinfo {author} {\bibfnamefont {A.}~\bibnamefont {Orb\'an}},
  \bibinfo {author} {\bibfnamefont {J.}~\bibnamefont {Werner}}, \bibinfo
  {author} {\bibfnamefont {S.}~\bibnamefont {Ros\'en}}, \bibinfo {author}
  {\bibfnamefont {R.~D.}\ \bibnamefont {Thomas}}, \bibinfo {author}
  {\bibfnamefont {I.}~\bibnamefont {Kashperka}}, \bibinfo {author}
  {\bibfnamefont {H.~A.~B.}\ \bibnamefont {Johansson}}, \bibinfo {author}
  {\bibfnamefont {D.}~\bibnamefont {Misra}}, \bibinfo {author} {\bibfnamefont
  {L.}~\bibnamefont {Br\"annholm}}, \bibinfo {author} {\bibfnamefont
  {M.}~\bibnamefont {Bj\"orkhage}}, \bibinfo {author} {\bibfnamefont
  {H.}~\bibnamefont {Cederquist}}, \ and\ \bibinfo {author} {\bibfnamefont
  {H.~T.}\ \bibnamefont {Schmidt}},\ }\href@noop {} {\bibfield  {journal}
  {\bibinfo  {journal} {Phys. Rev. Lett.}\ }\textbf {\bibinfo {volume} {103}},\
  \bibinfo {pages} {213002} (\bibinfo {year} {2009})}\BibitemShut {NoStop}%
\bibitem [{\citenamefont {Lange}\ \emph {et~al.}(2010)\citenamefont {Lange},
  \citenamefont {Froese}, \citenamefont {Menk}, \citenamefont {Varju},
  \citenamefont {Bastert}, \citenamefont {Blaum}, \citenamefont
  {L{\'o}pez-Urrutia}, \citenamefont {Fellenberger}, \citenamefont {Grieser},
  \citenamefont {von Hahn}, \citenamefont {Heber}, \citenamefont {K{\"u}hnel},
  \citenamefont {Laux}, \citenamefont {Orlov}, \citenamefont {Rappaport},
  \citenamefont {Repnow}, \citenamefont {Schr{\"o}ter}, \citenamefont
  {Schwalm}, \citenamefont {Shornikov}, \citenamefont {Sieber}, \citenamefont
  {Toker}, \citenamefont {Ullrich}, \citenamefont {Wolf},\ and\ \citenamefont
  {Zajfman}}]{Lange2010}%
  \BibitemOpen
  \bibfield  {author} {\bibinfo {author} {\bibfnamefont {M.}~\bibnamefont
  {Lange}}, \bibinfo {author} {\bibfnamefont {M.}~\bibnamefont {Froese}},
  \bibinfo {author} {\bibfnamefont {S.}~\bibnamefont {Menk}}, \bibinfo {author}
  {\bibfnamefont {J.}~\bibnamefont {Varju}}, \bibinfo {author} {\bibfnamefont
  {R.}~\bibnamefont {Bastert}}, \bibinfo {author} {\bibfnamefont
  {K.}~\bibnamefont {Blaum}}, \bibinfo {author} {\bibfnamefont {J.~R.~C.}\
  \bibnamefont {L{\'o}pez-Urrutia}}, \bibinfo {author} {\bibfnamefont
  {F.}~\bibnamefont {Fellenberger}}, \bibinfo {author} {\bibfnamefont
  {M.}~\bibnamefont {Grieser}}, \bibinfo {author} {\bibfnamefont
  {R.}~\bibnamefont {von Hahn}}, \bibinfo {author} {\bibfnamefont
  {O.}~\bibnamefont {Heber}}, \bibinfo {author} {\bibfnamefont {K.-U.}\
  \bibnamefont {K{\"u}hnel}}, \bibinfo {author} {\bibfnamefont
  {F.}~\bibnamefont {Laux}}, \bibinfo {author} {\bibfnamefont {D.~A.}\
  \bibnamefont {Orlov}}, \bibinfo {author} {\bibfnamefont {M.~L.}\ \bibnamefont
  {Rappaport}}, \bibinfo {author} {\bibfnamefont {R.}~\bibnamefont {Repnow}},
  \bibinfo {author} {\bibfnamefont {C.~D.}\ \bibnamefont {Schr{\"o}ter}},
  \bibinfo {author} {\bibfnamefont {D.}~\bibnamefont {Schwalm}}, \bibinfo
  {author} {\bibfnamefont {A.}~\bibnamefont {Shornikov}}, \bibinfo {author}
  {\bibfnamefont {T.}~\bibnamefont {Sieber}}, \bibinfo {author} {\bibfnamefont
  {Y.}~\bibnamefont {Toker}}, \bibinfo {author} {\bibfnamefont
  {J.}~\bibnamefont {Ullrich}}, \bibinfo {author} {\bibfnamefont
  {A.}~\bibnamefont {Wolf}}, \ and\ \bibinfo {author} {\bibfnamefont
  {D.}~\bibnamefont {Zajfman}},\ }\href@noop {} {\bibfield  {journal} {\bibinfo
   {journal} {Review of Scientific Instruments}\ }\textbf {\bibinfo {volume}
  {81}},\ \bibinfo {pages} {055105} (\bibinfo {year} {2010})}\BibitemShut
  {NoStop}%
\bibitem [{\citenamefont {von Hahn}\ \emph {et~al.}(2011)\citenamefont {von
  Hahn}, \citenamefont {Berg}, \citenamefont {Blaum}, \citenamefont
  {Lopez-Urrutia}, \citenamefont {Fellenberger}, \citenamefont {Froese},
  \citenamefont {Grieser}, \citenamefont {Krantz}, \citenamefont {K{\"u}hnel},
  \citenamefont {Lange}, \citenamefont {Menk}, \citenamefont {Laux},
  \citenamefont {Orlov}, \citenamefont {Repnow}, \citenamefont {Schr{\"o}ter},
  \citenamefont {Shornikov}, \citenamefont {Sieber}, \citenamefont {Ullrich},
  \citenamefont {Wolf}, \citenamefont {Rappaport},\ and\ \citenamefont
  {Zajfman}}]{Hahn2011}%
  \BibitemOpen
  \bibfield  {author} {\bibinfo {author} {\bibfnamefont {R.}~\bibnamefont {von
  Hahn}}, \bibinfo {author} {\bibfnamefont {F.}~\bibnamefont {Berg}}, \bibinfo
  {author} {\bibfnamefont {K.}~\bibnamefont {Blaum}}, \bibinfo {author}
  {\bibfnamefont {J.~C.}\ \bibnamefont {Lopez-Urrutia}}, \bibinfo {author}
  {\bibfnamefont {F.}~\bibnamefont {Fellenberger}}, \bibinfo {author}
  {\bibfnamefont {M.}~\bibnamefont {Froese}}, \bibinfo {author} {\bibfnamefont
  {M.}~\bibnamefont {Grieser}}, \bibinfo {author} {\bibfnamefont
  {C.}~\bibnamefont {Krantz}}, \bibinfo {author} {\bibfnamefont {K.-U.}\
  \bibnamefont {K{\"u}hnel}}, \bibinfo {author} {\bibfnamefont
  {M.}~\bibnamefont {Lange}}, \bibinfo {author} {\bibfnamefont
  {S.}~\bibnamefont {Menk}}, \bibinfo {author} {\bibfnamefont {F.}~\bibnamefont
  {Laux}}, \bibinfo {author} {\bibfnamefont {D.}~\bibnamefont {Orlov}},
  \bibinfo {author} {\bibfnamefont {R.}~\bibnamefont {Repnow}}, \bibinfo
  {author} {\bibfnamefont {C.}~\bibnamefont {Schr{\"o}ter}}, \bibinfo {author}
  {\bibfnamefont {A.}~\bibnamefont {Shornikov}}, \bibinfo {author}
  {\bibfnamefont {T.}~\bibnamefont {Sieber}}, \bibinfo {author} {\bibfnamefont
  {J.}~\bibnamefont {Ullrich}}, \bibinfo {author} {\bibfnamefont
  {A.}~\bibnamefont {Wolf}}, \bibinfo {author} {\bibfnamefont {M.}~\bibnamefont
  {Rappaport}}, \ and\ \bibinfo {author} {\bibfnamefont {D.}~\bibnamefont
  {Zajfman}},\ }\href@noop {} {\bibfield  {journal} {\bibinfo  {journal}
  {Nuclear Instruments and Methods in Physics Research Section B: Beam
  Interactions with Materials and Atoms}\ }\textbf {\bibinfo {volume} {269}},\
  \bibinfo {pages} {2871 } (\bibinfo {year} {2011})}\BibitemShut {NoStop}%
\bibitem [{\citenamefont {Nakano}\ \emph {et~al.}(2017)\citenamefont {Nakano},
  \citenamefont {Enomoto}, \citenamefont {Masunaga}, \citenamefont {Menk},
  \citenamefont {Bertier},\ and\ \citenamefont {Azuma}}]{Nakano2017}%
  \BibitemOpen
  \bibfield  {author} {\bibinfo {author} {\bibfnamefont {Y.}~\bibnamefont
  {Nakano}}, \bibinfo {author} {\bibfnamefont {Y.}~\bibnamefont {Enomoto}},
  \bibinfo {author} {\bibfnamefont {T.}~\bibnamefont {Masunaga}}, \bibinfo
  {author} {\bibfnamefont {S.}~\bibnamefont {Menk}}, \bibinfo {author}
  {\bibfnamefont {P.}~\bibnamefont {Bertier}}, \ and\ \bibinfo {author}
  {\bibfnamefont {T.}~\bibnamefont {Azuma}},\ }\href@noop {} {\bibfield
  {journal} {\bibinfo  {journal} {Review of Scientific Instruments}\ }\textbf
  {\bibinfo {volume} {88}},\ \bibinfo {pages} {033110} (\bibinfo {year}
  {2017})}\BibitemShut {NoStop}%
\bibitem [{\citenamefont {M{\o}ller}(1997)}]{Moller1997}%
  \BibitemOpen
  \bibfield  {author} {\bibinfo {author} {\bibfnamefont {S.~P.}\ \bibnamefont
  {M{\o}ller}},\ }\href@noop {} {\bibfield  {journal} {\bibinfo  {journal}
  {Nuclear Instruments and Methods in Physics Research Section A: Accelerators,
  Spectrometers, Detectors and Associated Equipment}\ }\textbf {\bibinfo
  {volume} {394}},\ \bibinfo {pages} {281 } (\bibinfo {year}
  {1997})}\BibitemShut {NoStop}%
\bibitem [{\citenamefont {von Hahn}\ \emph {et~al.}(2016)\citenamefont {von
  Hahn}, \citenamefont {Becker}, \citenamefont {Berg}, \citenamefont {Blaum},
  \citenamefont {Breitenfeldt}, \citenamefont {Fadil}, \citenamefont
  {Fellenberger}, \citenamefont {Froese}, \citenamefont {George}, \citenamefont
  {G{\"o}ck}, \citenamefont {Grieser}, \citenamefont {Grussie}, \citenamefont
  {Guerin}, \citenamefont {Heber}, \citenamefont {Herwig}, \citenamefont
  {Karthein}, \citenamefont {Krantz}, \citenamefont {Kreckel}, \citenamefont
  {Lange}, \citenamefont {Laux}, \citenamefont {Lohmann}, \citenamefont {Menk},
  \citenamefont {Meyer}, \citenamefont {Mishra}, \citenamefont {Novotn{\'y}},
  \citenamefont {O'Connor}, \citenamefont {Orlov}, \citenamefont {Rappaport},
  \citenamefont {Repnow}, \citenamefont {Saurabh}, \citenamefont {Schippers},
  \citenamefont {Schr{\"o}ter}, \citenamefont {Schwalm}, \citenamefont
  {Schweikhard}, \citenamefont {Sieber}, \citenamefont {Shornikov},
  \citenamefont {Spruck}, \citenamefont {Kumar}, \citenamefont {Ullrich},
  \citenamefont {Urbain}, \citenamefont {Vogel}, \citenamefont {Wilhelm},
  \citenamefont {Wolf},\ and\ \citenamefont {Zajfman}}]{Hahn2016}%
  \BibitemOpen
  \bibfield  {author} {\bibinfo {author} {\bibfnamefont {R.}~\bibnamefont {von
  Hahn}}, \bibinfo {author} {\bibfnamefont {A.}~\bibnamefont {Becker}},
  \bibinfo {author} {\bibfnamefont {F.}~\bibnamefont {Berg}}, \bibinfo {author}
  {\bibfnamefont {K.}~\bibnamefont {Blaum}}, \bibinfo {author} {\bibfnamefont
  {C.}~\bibnamefont {Breitenfeldt}}, \bibinfo {author} {\bibfnamefont
  {H.}~\bibnamefont {Fadil}}, \bibinfo {author} {\bibfnamefont
  {F.}~\bibnamefont {Fellenberger}}, \bibinfo {author} {\bibfnamefont
  {M.}~\bibnamefont {Froese}}, \bibinfo {author} {\bibfnamefont
  {S.}~\bibnamefont {George}}, \bibinfo {author} {\bibfnamefont
  {J.}~\bibnamefont {G{\"o}ck}}, \bibinfo {author} {\bibfnamefont
  {M.}~\bibnamefont {Grieser}}, \bibinfo {author} {\bibfnamefont
  {F.}~\bibnamefont {Grussie}}, \bibinfo {author} {\bibfnamefont {E.~A.}\
  \bibnamefont {Guerin}}, \bibinfo {author} {\bibfnamefont {O.}~\bibnamefont
  {Heber}}, \bibinfo {author} {\bibfnamefont {P.}~\bibnamefont {Herwig}},
  \bibinfo {author} {\bibfnamefont {J.}~\bibnamefont {Karthein}}, \bibinfo
  {author} {\bibfnamefont {C.}~\bibnamefont {Krantz}}, \bibinfo {author}
  {\bibfnamefont {H.}~\bibnamefont {Kreckel}}, \bibinfo {author} {\bibfnamefont
  {M.}~\bibnamefont {Lange}}, \bibinfo {author} {\bibfnamefont
  {F.}~\bibnamefont {Laux}}, \bibinfo {author} {\bibfnamefont {S.}~\bibnamefont
  {Lohmann}}, \bibinfo {author} {\bibfnamefont {S.}~\bibnamefont {Menk}},
  \bibinfo {author} {\bibfnamefont {C.}~\bibnamefont {Meyer}}, \bibinfo
  {author} {\bibfnamefont {P.~M.}\ \bibnamefont {Mishra}}, \bibinfo {author}
  {\bibfnamefont {O.}~\bibnamefont {Novotn{\'y}}}, \bibinfo {author}
  {\bibfnamefont {A.~P.}\ \bibnamefont {O'Connor}}, \bibinfo {author}
  {\bibfnamefont {D.~A.}\ \bibnamefont {Orlov}}, \bibinfo {author}
  {\bibfnamefont {M.~L.}\ \bibnamefont {Rappaport}}, \bibinfo {author}
  {\bibfnamefont {R.}~\bibnamefont {Repnow}}, \bibinfo {author} {\bibfnamefont
  {S.}~\bibnamefont {Saurabh}}, \bibinfo {author} {\bibfnamefont
  {S.}~\bibnamefont {Schippers}}, \bibinfo {author} {\bibfnamefont {C.~D.}\
  \bibnamefont {Schr{\"o}ter}}, \bibinfo {author} {\bibfnamefont
  {D.}~\bibnamefont {Schwalm}}, \bibinfo {author} {\bibfnamefont
  {L.}~\bibnamefont {Schweikhard}}, \bibinfo {author} {\bibfnamefont
  {T.}~\bibnamefont {Sieber}}, \bibinfo {author} {\bibfnamefont
  {A.}~\bibnamefont {Shornikov}}, \bibinfo {author} {\bibfnamefont
  {K.}~\bibnamefont {Spruck}}, \bibinfo {author} {\bibfnamefont {S.~S.}\
  \bibnamefont {Kumar}}, \bibinfo {author} {\bibfnamefont {J.}~\bibnamefont
  {Ullrich}}, \bibinfo {author} {\bibfnamefont {X.}~\bibnamefont {Urbain}},
  \bibinfo {author} {\bibfnamefont {S.}~\bibnamefont {Vogel}}, \bibinfo
  {author} {\bibfnamefont {P.}~\bibnamefont {Wilhelm}}, \bibinfo {author}
  {\bibfnamefont {A.}~\bibnamefont {Wolf}}, \ and\ \bibinfo {author}
  {\bibfnamefont {D.}~\bibnamefont {Zajfman}},\ }\href@noop {} {\bibfield
  {journal} {\bibinfo  {journal} {Review of Scientific Instruments}\ }\textbf
  {\bibinfo {volume} {87}},\ \bibinfo {pages} {063115} (\bibinfo {year}
  {2016})}\BibitemShut {NoStop}%
\bibitem [{\citenamefont {B\"ackstr\"om}\ \emph {et~al.}(2015)\citenamefont
  {B\"ackstr\"om}, \citenamefont {Hanstorp}, \citenamefont {Hole},
  \citenamefont {Kaminska}, \citenamefont {Nascimento}, \citenamefont {Blom},
  \citenamefont {Bj\"orkhage}, \citenamefont {K\"allberg}, \citenamefont
  {L\"ofgren}, \citenamefont {Reinhed}, \citenamefont {Ros\'en}, \citenamefont
  {Simonsson}, \citenamefont {Thomas}, \citenamefont {Mannervik}, \citenamefont
  {Schmidt},\ and\ \citenamefont {Cederquist}}]{Backstrom2015}%
  \BibitemOpen
  \bibfield  {author} {\bibinfo {author} {\bibfnamefont {E.}~\bibnamefont
  {B\"ackstr\"om}}, \bibinfo {author} {\bibfnamefont {D.}~\bibnamefont
  {Hanstorp}}, \bibinfo {author} {\bibfnamefont {O.~M.}\ \bibnamefont {Hole}},
  \bibinfo {author} {\bibfnamefont {M.}~\bibnamefont {Kaminska}}, \bibinfo
  {author} {\bibfnamefont {R.~F.}\ \bibnamefont {Nascimento}}, \bibinfo
  {author} {\bibfnamefont {M.}~\bibnamefont {Blom}}, \bibinfo {author}
  {\bibfnamefont {M.}~\bibnamefont {Bj\"orkhage}}, \bibinfo {author}
  {\bibfnamefont {A.}~\bibnamefont {K\"allberg}}, \bibinfo {author}
  {\bibfnamefont {P.}~\bibnamefont {L\"ofgren}}, \bibinfo {author}
  {\bibfnamefont {P.}~\bibnamefont {Reinhed}}, \bibinfo {author} {\bibfnamefont
  {S.}~\bibnamefont {Ros\'en}}, \bibinfo {author} {\bibfnamefont
  {A.}~\bibnamefont {Simonsson}}, \bibinfo {author} {\bibfnamefont {R.~D.}\
  \bibnamefont {Thomas}}, \bibinfo {author} {\bibfnamefont {S.}~\bibnamefont
  {Mannervik}}, \bibinfo {author} {\bibfnamefont {H.~T.}\ \bibnamefont
  {Schmidt}}, \ and\ \bibinfo {author} {\bibfnamefont {H.}~\bibnamefont
  {Cederquist}},\ }\href@noop {} {\bibfield  {journal} {\bibinfo  {journal}
  {Phys. Rev. Lett.}\ }\textbf {\bibinfo {volume} {114}},\ \bibinfo {pages}
  {143003} (\bibinfo {year} {2015})}\BibitemShut {NoStop}%
\bibitem [{\citenamefont {Schmidt}\ \emph {et~al.}(2017)\citenamefont
  {Schmidt}, \citenamefont {Eklund}, \citenamefont {Chartkunchand},
  \citenamefont {Anderson}, \citenamefont {Kami\ifmmode~\acute{n}\else
  \'{n}\fi{}ska}, \citenamefont {de~Ruette}, \citenamefont {Thomas},
  \citenamefont {Kristiansson}, \citenamefont {Gatchell}, \citenamefont
  {Reinhed}, \citenamefont {Ros\'en}, \citenamefont {Simonsson}, \citenamefont
  {K\"allberg}, \citenamefont {L\"ofgren}, \citenamefont {Mannervik},
  \citenamefont {Zettergren},\ and\ \citenamefont {Cederquist}}]{Schmidt2017}%
  \BibitemOpen
  \bibfield  {author} {\bibinfo {author} {\bibfnamefont {H.~T.}\ \bibnamefont
  {Schmidt}}, \bibinfo {author} {\bibfnamefont {G.}~\bibnamefont {Eklund}},
  \bibinfo {author} {\bibfnamefont {K.~C.}\ \bibnamefont {Chartkunchand}},
  \bibinfo {author} {\bibfnamefont {E.~K.}\ \bibnamefont {Anderson}}, \bibinfo
  {author} {\bibfnamefont {M.}~\bibnamefont {Kami\ifmmode~\acute{n}\else
  \'{n}\fi{}ska}}, \bibinfo {author} {\bibfnamefont {N.}~\bibnamefont
  {de~Ruette}}, \bibinfo {author} {\bibfnamefont {R.~D.}\ \bibnamefont
  {Thomas}}, \bibinfo {author} {\bibfnamefont {M.~K.}\ \bibnamefont
  {Kristiansson}}, \bibinfo {author} {\bibfnamefont {M.}~\bibnamefont
  {Gatchell}}, \bibinfo {author} {\bibfnamefont {P.}~\bibnamefont {Reinhed}},
  \bibinfo {author} {\bibfnamefont {S.}~\bibnamefont {Ros\'en}}, \bibinfo
  {author} {\bibfnamefont {A.}~\bibnamefont {Simonsson}}, \bibinfo {author}
  {\bibfnamefont {A.}~\bibnamefont {K\"allberg}}, \bibinfo {author}
  {\bibfnamefont {P.}~\bibnamefont {L\"ofgren}}, \bibinfo {author}
  {\bibfnamefont {S.}~\bibnamefont {Mannervik}}, \bibinfo {author}
  {\bibfnamefont {H.}~\bibnamefont {Zettergren}}, \ and\ \bibinfo {author}
  {\bibfnamefont {H.}~\bibnamefont {Cederquist}},\ }\href@noop {} {\bibfield
  {journal} {\bibinfo  {journal} {Phys. Rev. Lett.}\ }\textbf {\bibinfo
  {volume} {119}},\ \bibinfo {pages} {073001} (\bibinfo {year}
  {2017})}\BibitemShut {NoStop}%
\bibitem [{\citenamefont {Hansen}\ \emph {et~al.}(2001)\citenamefont {Hansen},
  \citenamefont {Andersen}, \citenamefont {Hvelplund}, \citenamefont
  {M\o{}ller}, \citenamefont {Pedersen},\ and\ \citenamefont
  {Petrunin}}]{Hansen2001}%
  \BibitemOpen
  \bibfield  {author} {\bibinfo {author} {\bibfnamefont {K.}~\bibnamefont
  {Hansen}}, \bibinfo {author} {\bibfnamefont {J.~U.}\ \bibnamefont
  {Andersen}}, \bibinfo {author} {\bibfnamefont {P.}~\bibnamefont {Hvelplund}},
  \bibinfo {author} {\bibfnamefont {S.~P.}\ \bibnamefont {M\o{}ller}}, \bibinfo
  {author} {\bibfnamefont {U.~V.}\ \bibnamefont {Pedersen}}, \ and\ \bibinfo
  {author} {\bibfnamefont {V.~V.}\ \bibnamefont {Petrunin}},\ }\href@noop {}
  {\bibfield  {journal} {\bibinfo  {journal} {Phys. Rev. Lett.}\ }\textbf
  {\bibinfo {volume} {87}},\ \bibinfo {pages} {123401} (\bibinfo {year}
  {2001})}\BibitemShut {NoStop}%
\bibitem [{\citenamefont {Kafle}\ \emph {et~al.}(2015)\citenamefont {Kafle},
  \citenamefont {Aviv}, \citenamefont {Chandrasekaran}, \citenamefont {Heber},
  \citenamefont {Rappaport}, \citenamefont {Rubinstein}, \citenamefont
  {Schwalm}, \citenamefont {Strasser},\ and\ \citenamefont
  {Zajfman}}]{Kafle2015}%
  \BibitemOpen
  \bibfield  {author} {\bibinfo {author} {\bibfnamefont {B.}~\bibnamefont
  {Kafle}}, \bibinfo {author} {\bibfnamefont {O.}~\bibnamefont {Aviv}},
  \bibinfo {author} {\bibfnamefont {V.}~\bibnamefont {Chandrasekaran}},
  \bibinfo {author} {\bibfnamefont {O.}~\bibnamefont {Heber}}, \bibinfo
  {author} {\bibfnamefont {M.~L.}\ \bibnamefont {Rappaport}}, \bibinfo {author}
  {\bibfnamefont {H.}~\bibnamefont {Rubinstein}}, \bibinfo {author}
  {\bibfnamefont {D.}~\bibnamefont {Schwalm}}, \bibinfo {author} {\bibfnamefont
  {D.}~\bibnamefont {Strasser}}, \ and\ \bibinfo {author} {\bibfnamefont
  {D.}~\bibnamefont {Zajfman}},\ }\href {\doibase 10.1103/PhysRevA.92.052503}
  {\bibfield  {journal} {\bibinfo  {journal} {Phys. Rev. A}\ }\textbf {\bibinfo
  {volume} {92}},\ \bibinfo {pages} {052503} (\bibinfo {year}
  {2015})}\BibitemShut {NoStop}%
\bibitem [{\citenamefont {Toker}\ \emph {et~al.}(2007)\citenamefont {Toker},
  \citenamefont {Aviv}, \citenamefont {Eritt}, \citenamefont {Rappaport},
  \citenamefont {Heber}, \citenamefont {Schwalm},\ and\ \citenamefont
  {Zajfman}}]{Toker2007}%
  \BibitemOpen
  \bibfield  {author} {\bibinfo {author} {\bibfnamefont {Y.}~\bibnamefont
  {Toker}}, \bibinfo {author} {\bibfnamefont {O.}~\bibnamefont {Aviv}},
  \bibinfo {author} {\bibfnamefont {M.}~\bibnamefont {Eritt}}, \bibinfo
  {author} {\bibfnamefont {M.~L.}\ \bibnamefont {Rappaport}}, \bibinfo {author}
  {\bibfnamefont {O.}~\bibnamefont {Heber}}, \bibinfo {author} {\bibfnamefont
  {D.}~\bibnamefont {Schwalm}}, \ and\ \bibinfo {author} {\bibfnamefont
  {D.}~\bibnamefont {Zajfman}},\ }\href {\doibase 10.1103/PhysRevA.76.053201}
  {\bibfield  {journal} {\bibinfo  {journal} {Phys. Rev. A}\ }\textbf {\bibinfo
  {volume} {76}},\ \bibinfo {pages} {053201} (\bibinfo {year}
  {2007})}\BibitemShut {NoStop}%
\bibitem [{\citenamefont {Breitenfeldt}\ \emph {et~al.}(2016)\citenamefont
  {Breitenfeldt}, \citenamefont {Blaum}, \citenamefont {Froese}, \citenamefont
  {George}, \citenamefont {Guzm\'an-Ram\'{\i}rez}, \citenamefont {Lange},
  \citenamefont {Menk}, \citenamefont {Schweikhard},\ and\ \citenamefont
  {Wolf}}]{Breitenfeldt2016}%
  \BibitemOpen
  \bibfield  {author} {\bibinfo {author} {\bibfnamefont {C.}~\bibnamefont
  {Breitenfeldt}}, \bibinfo {author} {\bibfnamefont {K.}~\bibnamefont {Blaum}},
  \bibinfo {author} {\bibfnamefont {M.~W.}\ \bibnamefont {Froese}}, \bibinfo
  {author} {\bibfnamefont {S.}~\bibnamefont {George}}, \bibinfo {author}
  {\bibfnamefont {G.}~\bibnamefont {Guzm\'an-Ram\'{\i}rez}}, \bibinfo {author}
  {\bibfnamefont {M.}~\bibnamefont {Lange}}, \bibinfo {author} {\bibfnamefont
  {S.}~\bibnamefont {Menk}}, \bibinfo {author} {\bibfnamefont {L.}~\bibnamefont
  {Schweikhard}}, \ and\ \bibinfo {author} {\bibfnamefont {A.}~\bibnamefont
  {Wolf}},\ }\href {\doibase 10.1103/PhysRevA.94.033407} {\bibfield  {journal}
  {\bibinfo  {journal} {Phys. Rev. A}\ }\textbf {\bibinfo {volume} {94}},\
  \bibinfo {pages} {033407} (\bibinfo {year} {2016})}\BibitemShut {NoStop}%
\bibitem [{\citenamefont {Fedor}\ \emph {et~al.}(2005)\citenamefont {Fedor},
  \citenamefont {Hansen}, \citenamefont {Andersen},\ and\ \citenamefont
  {Hvelplund}}]{Fedor2005}%
  \BibitemOpen
  \bibfield  {author} {\bibinfo {author} {\bibfnamefont {J.}~\bibnamefont
  {Fedor}}, \bibinfo {author} {\bibfnamefont {K.}~\bibnamefont {Hansen}},
  \bibinfo {author} {\bibfnamefont {J.~U.}\ \bibnamefont {Andersen}}, \ and\
  \bibinfo {author} {\bibfnamefont {P.}~\bibnamefont {Hvelplund}},\ }\href
  {\doibase 10.1103/PhysRevLett.94.113201} {\bibfield  {journal} {\bibinfo
  {journal} {Phys. Rev. Lett.}\ }\textbf {\bibinfo {volume} {94}},\ \bibinfo
  {pages} {113201} (\bibinfo {year} {2005})}\BibitemShut {NoStop}%
\bibitem [{\citenamefont {Hansen}\ \emph {et~al.}(2017)\citenamefont {Hansen},
  \citenamefont {Stockett}, \citenamefont {Kaminska}, \citenamefont
  {Nascimento}, \citenamefont {Anderson}, \citenamefont {Gatchell},
  \citenamefont {Chartkunchand}, \citenamefont {Eklund}, \citenamefont
  {Zettergren}, \citenamefont {Schmidt},\ and\ \citenamefont
  {Cederquist}}]{Hansen2017}%
  \BibitemOpen
  \bibfield  {author} {\bibinfo {author} {\bibfnamefont {K.}~\bibnamefont
  {Hansen}}, \bibinfo {author} {\bibfnamefont {M.~H.}\ \bibnamefont
  {Stockett}}, \bibinfo {author} {\bibfnamefont {M.}~\bibnamefont {Kaminska}},
  \bibinfo {author} {\bibfnamefont {R.~F.}\ \bibnamefont {Nascimento}},
  \bibinfo {author} {\bibfnamefont {E.~K.}\ \bibnamefont {Anderson}}, \bibinfo
  {author} {\bibfnamefont {M.}~\bibnamefont {Gatchell}}, \bibinfo {author}
  {\bibfnamefont {K.~C.}\ \bibnamefont {Chartkunchand}}, \bibinfo {author}
  {\bibfnamefont {G.}~\bibnamefont {Eklund}}, \bibinfo {author} {\bibfnamefont
  {H.}~\bibnamefont {Zettergren}}, \bibinfo {author} {\bibfnamefont {H.~T.}\
  \bibnamefont {Schmidt}}, \ and\ \bibinfo {author} {\bibfnamefont
  {H.}~\bibnamefont {Cederquist}},\ }\href {\doibase
  10.1103/PhysRevA.95.022511} {\bibfield  {journal} {\bibinfo  {journal} {Phys.
  Rev. A}\ }\textbf {\bibinfo {volume} {95}},\ \bibinfo {pages} {022511}
  (\bibinfo {year} {2017})}\BibitemShut {NoStop}%
\bibitem [{\citenamefont {Andersen}\ \emph {et~al.}(2001)\citenamefont
  {Andersen}, \citenamefont {Gottrup}, \citenamefont {Hansen}, \citenamefont
  {Hvelplund},\ and\ \citenamefont {Larsson}}]{Andersen2001}%
  \BibitemOpen
  \bibfield  {author} {\bibinfo {author} {\bibfnamefont {J.~U.}\ \bibnamefont
  {Andersen}}, \bibinfo {author} {\bibfnamefont {C.}~\bibnamefont {Gottrup}},
  \bibinfo {author} {\bibfnamefont {K.}~\bibnamefont {Hansen}}, \bibinfo
  {author} {\bibfnamefont {P.}~\bibnamefont {Hvelplund}}, \ and\ \bibinfo
  {author} {\bibfnamefont {M.~O.}\ \bibnamefont {Larsson}},\ }\href {\doibase
  10.1007/s100530170022} {\bibfield  {journal} {\bibinfo  {journal} {Eur. Phys.
  J. D}\ }\textbf {\bibinfo {volume} {17}},\ \bibinfo {pages} {189} (\bibinfo
  {year} {2001})}\BibitemShut {NoStop}%
\bibitem [{\citenamefont {Tomita}\ \emph {et~al.}(2006)\citenamefont {Tomita},
  \citenamefont {Andersen}, \citenamefont {Cederquist}, \citenamefont
  {Concina}, \citenamefont {Echt}, \citenamefont {Forster}, \citenamefont
  {Hansen}, \citenamefont {Huber}, \citenamefont {Hvelplund}, \citenamefont
  {Jensen}, \citenamefont {Liu}, \citenamefont {Manil}, \citenamefont
  {Maunoury}, \citenamefont {Nielsen}, \citenamefont {Rangama}, \citenamefont
  {Schmidt},\ and\ \citenamefont {Zettergren}}]{Tomita2006}%
  \BibitemOpen
  \bibfield  {author} {\bibinfo {author} {\bibfnamefont {S.}~\bibnamefont
  {Tomita}}, \bibinfo {author} {\bibfnamefont {J.~U.}\ \bibnamefont
  {Andersen}}, \bibinfo {author} {\bibfnamefont {H.}~\bibnamefont
  {Cederquist}}, \bibinfo {author} {\bibfnamefont {B.}~\bibnamefont {Concina}},
  \bibinfo {author} {\bibfnamefont {O.}~\bibnamefont {Echt}}, \bibinfo {author}
  {\bibfnamefont {J.~S.}\ \bibnamefont {Forster}}, \bibinfo {author}
  {\bibfnamefont {K.}~\bibnamefont {Hansen}}, \bibinfo {author} {\bibfnamefont
  {B.~A.}\ \bibnamefont {Huber}}, \bibinfo {author} {\bibfnamefont
  {P.}~\bibnamefont {Hvelplund}}, \bibinfo {author} {\bibfnamefont
  {J.}~\bibnamefont {Jensen}}, \bibinfo {author} {\bibfnamefont
  {B.}~\bibnamefont {Liu}}, \bibinfo {author} {\bibfnamefont {B.}~\bibnamefont
  {Manil}}, \bibinfo {author} {\bibfnamefont {L.}~\bibnamefont {Maunoury}},
  \bibinfo {author} {\bibfnamefont {S.~B.}\ \bibnamefont {Nielsen}}, \bibinfo
  {author} {\bibfnamefont {J.}~\bibnamefont {Rangama}}, \bibinfo {author}
  {\bibfnamefont {H.~T.}\ \bibnamefont {Schmidt}}, \ and\ \bibinfo {author}
  {\bibfnamefont {H.}~\bibnamefont {Zettergren}},\ }\href@noop {} {\bibfield
  {journal} {\bibinfo  {journal} {The Journal of Chemical Physics}\ }\textbf
  {\bibinfo {volume} {124}},\ \bibinfo {pages} {024310} (\bibinfo {year}
  {2006})}\BibitemShut {NoStop}%
\bibitem [{\citenamefont {Goto}\ \emph {et~al.}(2013)\citenamefont {Goto},
  \citenamefont {Sund{\'e}n}, \citenamefont {Shiromaru}, \citenamefont
  {Matsumoto}, \citenamefont {Tanuma}, \citenamefont {Azuma},\ and\
  \citenamefont {Hansen}}]{Goto2013}%
  \BibitemOpen
  \bibfield  {author} {\bibinfo {author} {\bibfnamefont {M.}~\bibnamefont
  {Goto}}, \bibinfo {author} {\bibfnamefont {A.~E.~K.}\ \bibnamefont
  {Sund{\'e}n}}, \bibinfo {author} {\bibfnamefont {H.}~\bibnamefont
  {Shiromaru}}, \bibinfo {author} {\bibfnamefont {J.}~\bibnamefont
  {Matsumoto}}, \bibinfo {author} {\bibfnamefont {H.}~\bibnamefont {Tanuma}},
  \bibinfo {author} {\bibfnamefont {T.}~\bibnamefont {Azuma}}, \ and\ \bibinfo
  {author} {\bibfnamefont {K.}~\bibnamefont {Hansen}},\ }\href@noop {}
  {\bibfield  {journal} {\bibinfo  {journal} {The Journal of Chemical Physics}\
  }\textbf {\bibinfo {volume} {139}},\ \bibinfo {pages} {054306} (\bibinfo
  {year} {2013})}\BibitemShut {NoStop}%
\bibitem [{\citenamefont {Shiromaru}\ \emph {et~al.}(2015)\citenamefont
  {Shiromaru}, \citenamefont {Furukawa}, \citenamefont {Ito}, \citenamefont
  {Kono}, \citenamefont {Tanuma}, \citenamefont {Matsumoto}, \citenamefont
  {Goto}, \citenamefont {Majima}, \citenamefont {Sund{\'e}n}, \citenamefont
  {Najafian}, \citenamefont {Pettersson}, \citenamefont {Dynefors},
  \citenamefont {Hansen},\ and\ \citenamefont {Azuma}}]{Shiromaru2015}%
  \BibitemOpen
  \bibfield  {author} {\bibinfo {author} {\bibfnamefont {H.}~\bibnamefont
  {Shiromaru}}, \bibinfo {author} {\bibfnamefont {T.}~\bibnamefont {Furukawa}},
  \bibinfo {author} {\bibfnamefont {G.}~\bibnamefont {Ito}}, \bibinfo {author}
  {\bibfnamefont {N.}~\bibnamefont {Kono}}, \bibinfo {author} {\bibfnamefont
  {H.}~\bibnamefont {Tanuma}}, \bibinfo {author} {\bibfnamefont
  {J.}~\bibnamefont {Matsumoto}}, \bibinfo {author} {\bibfnamefont
  {M.}~\bibnamefont {Goto}}, \bibinfo {author} {\bibfnamefont {T.}~\bibnamefont
  {Majima}}, \bibinfo {author} {\bibfnamefont {A.~E.~K.}\ \bibnamefont
  {Sund{\'e}n}}, \bibinfo {author} {\bibfnamefont {K.}~\bibnamefont
  {Najafian}}, \bibinfo {author} {\bibfnamefont {M.~S.}\ \bibnamefont
  {Pettersson}}, \bibinfo {author} {\bibfnamefont {B.}~\bibnamefont
  {Dynefors}}, \bibinfo {author} {\bibfnamefont {K.}~\bibnamefont {Hansen}}, \
  and\ \bibinfo {author} {\bibfnamefont {T.}~\bibnamefont {Azuma}},\ }\href
  {http://stacks.iop.org/1742-6596/635/i=1/a=012035} {\bibfield  {journal}
  {\bibinfo  {journal} {Journal of Physics: Conference Series}\ }\textbf
  {\bibinfo {volume} {635}},\ \bibinfo {pages} {012035} (\bibinfo {year}
  {2015})}\BibitemShut {NoStop}%
\bibitem [{\citenamefont {Najafian}\ \emph {et~al.}(2014)\citenamefont
  {Najafian}, \citenamefont {Pettersson}, \citenamefont {Dynefors},
  \citenamefont {Shiromaru}, \citenamefont {Matsumoto}, \citenamefont {Tanuma},
  \citenamefont {Furukawa}, \citenamefont {Azuma},\ and\ \citenamefont
  {Hansen}}]{Najafian2014}%
  \BibitemOpen
  \bibfield  {author} {\bibinfo {author} {\bibfnamefont {K.}~\bibnamefont
  {Najafian}}, \bibinfo {author} {\bibfnamefont {M.~S.}\ \bibnamefont
  {Pettersson}}, \bibinfo {author} {\bibfnamefont {B.}~\bibnamefont
  {Dynefors}}, \bibinfo {author} {\bibfnamefont {H.}~\bibnamefont {Shiromaru}},
  \bibinfo {author} {\bibfnamefont {J.}~\bibnamefont {Matsumoto}}, \bibinfo
  {author} {\bibfnamefont {H.}~\bibnamefont {Tanuma}}, \bibinfo {author}
  {\bibfnamefont {T.}~\bibnamefont {Furukawa}}, \bibinfo {author}
  {\bibfnamefont {T.}~\bibnamefont {Azuma}}, \ and\ \bibinfo {author}
  {\bibfnamefont {K.}~\bibnamefont {Hansen}},\ }\href@noop {} {\bibfield
  {journal} {\bibinfo  {journal} {The Journal of Chemical Physics}\ }\textbf
  {\bibinfo {volume} {140}},\ \bibinfo {pages} {104311} (\bibinfo {year}
  {2014})}\BibitemShut {NoStop}%
\bibitem [{\citenamefont {Andersen}\ \emph {et~al.}(2003)\citenamefont
  {Andersen}, \citenamefont {Cederquist}, \citenamefont {Forster},
  \citenamefont {Huber}, \citenamefont {Hvelplund}, \citenamefont {Jensen},
  \citenamefont {Liu}, \citenamefont {Manil}, \citenamefont {Maunoury},
  \citenamefont {Br{\o}ndsted~Nielsen}, \citenamefont {Pedersen}, \citenamefont
  {Schmidt}, \citenamefont {Tomita},\ and\ \citenamefont
  {Zettergren}}]{Andersen2003_b}%
  \BibitemOpen
  \bibfield  {author} {\bibinfo {author} {\bibfnamefont {J.~U.}\ \bibnamefont
  {Andersen}}, \bibinfo {author} {\bibfnamefont {H.}~\bibnamefont
  {Cederquist}}, \bibinfo {author} {\bibfnamefont {J.~S.}\ \bibnamefont
  {Forster}}, \bibinfo {author} {\bibfnamefont {B.~A.}\ \bibnamefont {Huber}},
  \bibinfo {author} {\bibfnamefont {P.}~\bibnamefont {Hvelplund}}, \bibinfo
  {author} {\bibfnamefont {J.}~\bibnamefont {Jensen}}, \bibinfo {author}
  {\bibfnamefont {B.}~\bibnamefont {Liu}}, \bibinfo {author} {\bibfnamefont
  {B.}~\bibnamefont {Manil}}, \bibinfo {author} {\bibfnamefont
  {L.}~\bibnamefont {Maunoury}}, \bibinfo {author} {\bibfnamefont
  {S.}~\bibnamefont {Br{\o}ndsted~Nielsen}}, \bibinfo {author} {\bibfnamefont
  {U.~V.}\ \bibnamefont {Pedersen}}, \bibinfo {author} {\bibfnamefont {H.~T.}\
  \bibnamefont {Schmidt}}, \bibinfo {author} {\bibfnamefont {S.}~\bibnamefont
  {Tomita}}, \ and\ \bibinfo {author} {\bibfnamefont {H.}~\bibnamefont
  {Zettergren}},\ }\href {\doibase 10.1140/epjd/e2003-00093-9} {\bibfield
  {journal} {\bibinfo  {journal} {The European Physical Journal D - Atomic,
  Molecular, Optical and Plasma Physics}\ }\textbf {\bibinfo {volume} {25}},\
  \bibinfo {pages} {139} (\bibinfo {year} {2003})}\BibitemShut {NoStop}%
\bibitem [{\citenamefont {Andersen}\ \emph {et~al.}(2004)\citenamefont
  {Andersen}, \citenamefont {Cederquist}, \citenamefont {Forster},
  \citenamefont {Huber}, \citenamefont {Hvelplund}, \citenamefont {Jensen},
  \citenamefont {Liu}, \citenamefont {Manil}, \citenamefont {Maunoury},
  \citenamefont {Brondsted~Nielsen}, \citenamefont {Pedersen}, \citenamefont
  {Rangama}, \citenamefont {Schmidt}, \citenamefont {Tomita},\ and\
  \citenamefont {Zettergren}}]{Andersen2004}%
  \BibitemOpen
  \bibfield  {author} {\bibinfo {author} {\bibfnamefont {J.~U.}\ \bibnamefont
  {Andersen}}, \bibinfo {author} {\bibfnamefont {H.}~\bibnamefont
  {Cederquist}}, \bibinfo {author} {\bibfnamefont {J.~S.}\ \bibnamefont
  {Forster}}, \bibinfo {author} {\bibfnamefont {B.~A.}\ \bibnamefont {Huber}},
  \bibinfo {author} {\bibfnamefont {P.}~\bibnamefont {Hvelplund}}, \bibinfo
  {author} {\bibfnamefont {J.}~\bibnamefont {Jensen}}, \bibinfo {author}
  {\bibfnamefont {B.}~\bibnamefont {Liu}}, \bibinfo {author} {\bibfnamefont
  {B.}~\bibnamefont {Manil}}, \bibinfo {author} {\bibfnamefont
  {L.}~\bibnamefont {Maunoury}}, \bibinfo {author} {\bibfnamefont
  {S.}~\bibnamefont {Brondsted~Nielsen}}, \bibinfo {author} {\bibfnamefont
  {U.~V.}\ \bibnamefont {Pedersen}}, \bibinfo {author} {\bibfnamefont
  {J.}~\bibnamefont {Rangama}}, \bibinfo {author} {\bibfnamefont {H.~T.}\
  \bibnamefont {Schmidt}}, \bibinfo {author} {\bibfnamefont {S.}~\bibnamefont
  {Tomita}}, \ and\ \bibinfo {author} {\bibfnamefont {H.}~\bibnamefont
  {Zettergren}},\ }\href {\doibase 10.1039/B316845J} {\bibfield  {journal}
  {\bibinfo  {journal} {Phys. Chem. Chem. Phys.}\ }\textbf {\bibinfo {volume}
  {6}},\ \bibinfo {pages} {2676} (\bibinfo {year} {2004})}\BibitemShut
  {NoStop}%
\bibitem [{\citenamefont {Nielsen}\ \emph {et~al.}(2004)\citenamefont
  {Nielsen}, \citenamefont {Andersen}, \citenamefont {Hvelplund}, \citenamefont
  {Liu},\ and\ \citenamefont {Tomita}}]{Nielsen2004}%
  \BibitemOpen
  \bibfield  {author} {\bibinfo {author} {\bibfnamefont {S.~B.}\ \bibnamefont
  {Nielsen}}, \bibinfo {author} {\bibfnamefont {J.~U.}\ \bibnamefont
  {Andersen}}, \bibinfo {author} {\bibfnamefont {P.}~\bibnamefont {Hvelplund}},
  \bibinfo {author} {\bibfnamefont {B.}~\bibnamefont {Liu}}, \ and\ \bibinfo
  {author} {\bibfnamefont {S.}~\bibnamefont {Tomita}},\ }\href
  {http://stacks.iop.org/0953-4075/37/i=8/a=R01} {\bibfield  {journal}
  {\bibinfo  {journal} {Journal of Physics B: Atomic, Molecular and Optical
  Physics}\ }\textbf {\bibinfo {volume} {37}},\ \bibinfo {pages} {R25}
  (\bibinfo {year} {2004})}\BibitemShut {NoStop}%
\bibitem [{\citenamefont {Martin}\ \emph {et~al.}(2013)\citenamefont {Martin},
  \citenamefont {Bernard}, \citenamefont {Br\'edy}, \citenamefont {Concina},
  \citenamefont {Joblin}, \citenamefont {Ji}, \citenamefont {Ortega},\ and\
  \citenamefont {Chen}}]{Martin2013}%
  \BibitemOpen
  \bibfield  {author} {\bibinfo {author} {\bibfnamefont {S.}~\bibnamefont
  {Martin}}, \bibinfo {author} {\bibfnamefont {J.}~\bibnamefont {Bernard}},
  \bibinfo {author} {\bibfnamefont {R.}~\bibnamefont {Br\'edy}}, \bibinfo
  {author} {\bibfnamefont {B.}~\bibnamefont {Concina}}, \bibinfo {author}
  {\bibfnamefont {C.}~\bibnamefont {Joblin}}, \bibinfo {author} {\bibfnamefont
  {M.}~\bibnamefont {Ji}}, \bibinfo {author} {\bibfnamefont {C.}~\bibnamefont
  {Ortega}}, \ and\ \bibinfo {author} {\bibfnamefont {L.}~\bibnamefont
  {Chen}},\ }\href {\doibase 10.1103/PhysRevLett.110.063003} {\bibfield
  {journal} {\bibinfo  {journal} {Phys. Rev. Lett.}\ }\textbf {\bibinfo
  {volume} {110}},\ \bibinfo {pages} {063003} (\bibinfo {year}
  {2013})}\BibitemShut {NoStop}%
\bibitem [{\citenamefont {Ji}\ \emph {et~al.}(2013)\citenamefont {Ji},
  \citenamefont {Br{\'e}dy}, \citenamefont {Chen}, \citenamefont {Bernard},
  \citenamefont {Concina}, \citenamefont {Montagne}, \citenamefont {Cassimi},\
  and\ \citenamefont {Martin}}]{Ji2013}%
  \BibitemOpen
  \bibfield  {author} {\bibinfo {author} {\bibfnamefont {M.}~\bibnamefont
  {Ji}}, \bibinfo {author} {\bibfnamefont {R.}~\bibnamefont {Br{\'e}dy}},
  \bibinfo {author} {\bibfnamefont {L.}~\bibnamefont {Chen}}, \bibinfo {author}
  {\bibfnamefont {J.}~\bibnamefont {Bernard}}, \bibinfo {author} {\bibfnamefont
  {B.}~\bibnamefont {Concina}}, \bibinfo {author} {\bibfnamefont
  {G.}~\bibnamefont {Montagne}}, \bibinfo {author} {\bibfnamefont
  {A.}~\bibnamefont {Cassimi}}, \ and\ \bibinfo {author} {\bibfnamefont
  {S.}~\bibnamefont {Martin}},\ }\href@noop {} {\bibfield  {journal} {\bibinfo
  {journal} {Physica Scripta}\ }\textbf {\bibinfo {volume} {2013}},\ \bibinfo
  {pages} {014091} (\bibinfo {year} {2013})}\BibitemShut {NoStop}%
\bibitem [{\citenamefont {Rajput}\ \emph {et~al.}(2008)\citenamefont {Rajput},
  \citenamefont {Lammich},\ and\ \citenamefont {Andersen}}]{Rajput2008}%
  \BibitemOpen
  \bibfield  {author} {\bibinfo {author} {\bibfnamefont {J.}~\bibnamefont
  {Rajput}}, \bibinfo {author} {\bibfnamefont {L.}~\bibnamefont {Lammich}}, \
  and\ \bibinfo {author} {\bibfnamefont {L.~H.}\ \bibnamefont {Andersen}},\
  }\href {\doibase 10.1103/PhysRevLett.100.153001} {\bibfield  {journal}
  {\bibinfo  {journal} {Phys. Rev. Lett.}\ }\textbf {\bibinfo {volume} {100}},\
  \bibinfo {pages} {153001} (\bibinfo {year} {2008})}\BibitemShut {NoStop}%
\bibitem [{\citenamefont {Menk}\ \emph {et~al.}(2014)\citenamefont {Menk},
  \citenamefont {Das}, \citenamefont {Blaum}, \citenamefont {Froese},
  \citenamefont {Lange}, \citenamefont {Mukherjee}, \citenamefont {Repnow},
  \citenamefont {Schwalm}, \citenamefont {von Hahn},\ and\ \citenamefont
  {Wolf}}]{Menk2014}%
  \BibitemOpen
  \bibfield  {author} {\bibinfo {author} {\bibfnamefont {S.}~\bibnamefont
  {Menk}}, \bibinfo {author} {\bibfnamefont {S.}~\bibnamefont {Das}}, \bibinfo
  {author} {\bibfnamefont {K.}~\bibnamefont {Blaum}}, \bibinfo {author}
  {\bibfnamefont {M.~W.}\ \bibnamefont {Froese}}, \bibinfo {author}
  {\bibfnamefont {M.}~\bibnamefont {Lange}}, \bibinfo {author} {\bibfnamefont
  {M.}~\bibnamefont {Mukherjee}}, \bibinfo {author} {\bibfnamefont
  {R.}~\bibnamefont {Repnow}}, \bibinfo {author} {\bibfnamefont
  {D.}~\bibnamefont {Schwalm}}, \bibinfo {author} {\bibfnamefont
  {R.}~\bibnamefont {von Hahn}}, \ and\ \bibinfo {author} {\bibfnamefont
  {A.}~\bibnamefont {Wolf}},\ }\href {\doibase 10.1103/PhysRevA.89.022502}
  {\bibfield  {journal} {\bibinfo  {journal} {Phys. Rev. A}\ }\textbf {\bibinfo
  {volume} {89}},\ \bibinfo {pages} {022502} (\bibinfo {year}
  {2014})}\BibitemShut {NoStop}%
\bibitem [{\citenamefont {Aviv}\ \emph {et~al.}(2011)\citenamefont {Aviv},
  \citenamefont {Toker}, \citenamefont {Strasser}, \citenamefont {Rappaport},
  \citenamefont {Heber}, \citenamefont {Schwalm},\ and\ \citenamefont
  {Zajfman}}]{Aviv2011}%
  \BibitemOpen
  \bibfield  {author} {\bibinfo {author} {\bibfnamefont {O.}~\bibnamefont
  {Aviv}}, \bibinfo {author} {\bibfnamefont {Y.}~\bibnamefont {Toker}},
  \bibinfo {author} {\bibfnamefont {D.}~\bibnamefont {Strasser}}, \bibinfo
  {author} {\bibfnamefont {M.~L.}\ \bibnamefont {Rappaport}}, \bibinfo {author}
  {\bibfnamefont {O.}~\bibnamefont {Heber}}, \bibinfo {author} {\bibfnamefont
  {D.}~\bibnamefont {Schwalm}}, \ and\ \bibinfo {author} {\bibfnamefont
  {D.}~\bibnamefont {Zajfman}},\ }\href {\doibase 10.1103/PhysRevA.83.023201}
  {\bibfield  {journal} {\bibinfo  {journal} {Phys. Rev. A}\ }\textbf {\bibinfo
  {volume} {83}},\ \bibinfo {pages} {023201} (\bibinfo {year}
  {2011})}\BibitemShut {NoStop}%
\bibitem [{\citenamefont {National Electrostatic~Corp.}(2017)}]{SNICS}%
  \BibitemOpen
  \bibfield  {author} {\bibinfo {author} {\bibfnamefont {U.}~\bibnamefont
  {National Electrostatic~Corp.}},\ }\href@noop {} {\enquote {\bibinfo {title}
  {Source of negative ions by cesium sputtering - {SNICS II}},}\ } (\bibinfo
  {year} {2017})\BibitemShut {NoStop}%
\bibitem [{\citenamefont {Wucher}\ and\ \citenamefont
  {Garrison}(1996)}]{Wucher1996}%
  \BibitemOpen
  \bibfield  {author} {\bibinfo {author} {\bibfnamefont {A.}~\bibnamefont
  {Wucher}}\ and\ \bibinfo {author} {\bibfnamefont {B.~J.}\ \bibnamefont
  {Garrison}},\ }\href@noop {} {\bibfield  {journal} {\bibinfo  {journal} {The
  Journal of Chemical Physics}\ }\textbf {\bibinfo {volume} {105}},\ \bibinfo
  {pages} {5999} (\bibinfo {year} {1996})}\BibitemShut {NoStop}%
\bibitem [{\citenamefont {Bilodeau}\ \emph {et~al.}(1998)\citenamefont
  {Bilodeau}, \citenamefont {Scheer},\ and\ \citenamefont
  {Haugen}}]{Bilodeau1998}%
  \BibitemOpen
  \bibfield  {author} {\bibinfo {author} {\bibfnamefont {R.~C.}\ \bibnamefont
  {Bilodeau}}, \bibinfo {author} {\bibfnamefont {M.}~\bibnamefont {Scheer}}, \
  and\ \bibinfo {author} {\bibfnamefont {H.~K.}\ \bibnamefont {Haugen}},\
  }\href {http://stacks.iop.org/0953-4075/31/i=17/a=013} {\bibfield  {journal}
  {\bibinfo  {journal} {Journal of Physics B: Atomic, Molecular and Optical
  Physics}\ }\textbf {\bibinfo {volume} {31}},\ \bibinfo {pages} {3885}
  (\bibinfo {year} {1998})}\BibitemShut {NoStop}%
\bibitem [{\citenamefont {Froese}\ \emph {et~al.}(2011)\citenamefont {Froese},
  \citenamefont {Blaum}, \citenamefont {Fellenberger}, \citenamefont {Grieser},
  \citenamefont {Lange}, \citenamefont {Laux}, \citenamefont {Menk},
  \citenamefont {Orlov}, \citenamefont {Repnow}, \citenamefont {Sieber},
  \citenamefont {Toker}, \citenamefont {von Hahn},\ and\ \citenamefont
  {Wolf}}]{Froese2011}%
  \BibitemOpen
  \bibfield  {author} {\bibinfo {author} {\bibfnamefont {M.~W.}\ \bibnamefont
  {Froese}}, \bibinfo {author} {\bibfnamefont {K.}~\bibnamefont {Blaum}},
  \bibinfo {author} {\bibfnamefont {F.}~\bibnamefont {Fellenberger}}, \bibinfo
  {author} {\bibfnamefont {M.}~\bibnamefont {Grieser}}, \bibinfo {author}
  {\bibfnamefont {M.}~\bibnamefont {Lange}}, \bibinfo {author} {\bibfnamefont
  {F.}~\bibnamefont {Laux}}, \bibinfo {author} {\bibfnamefont {S.}~\bibnamefont
  {Menk}}, \bibinfo {author} {\bibfnamefont {D.~A.}\ \bibnamefont {Orlov}},
  \bibinfo {author} {\bibfnamefont {R.}~\bibnamefont {Repnow}}, \bibinfo
  {author} {\bibfnamefont {T.}~\bibnamefont {Sieber}}, \bibinfo {author}
  {\bibfnamefont {Y.}~\bibnamefont {Toker}}, \bibinfo {author} {\bibfnamefont
  {R.}~\bibnamefont {von Hahn}}, \ and\ \bibinfo {author} {\bibfnamefont
  {A.}~\bibnamefont {Wolf}},\ }\href {\doibase 10.1103/PhysRevA.83.023202}
  {\bibfield  {journal} {\bibinfo  {journal} {Phys. Rev. A}\ }\textbf {\bibinfo
  {volume} {83}},\ \bibinfo {pages} {023202} (\bibinfo {year}
  {2011})}\BibitemShut {NoStop}%
\bibitem [{\citenamefont {Weisskopf}(1937)}]{Weisskopf1937}%
  \BibitemOpen
  \bibfield  {author} {\bibinfo {author} {\bibfnamefont {V.}~\bibnamefont
  {Weisskopf}},\ }\href {\doibase 10.1103/PhysRev.52.295} {\bibfield  {journal}
  {\bibinfo  {journal} {Phys. Rev.}\ }\textbf {\bibinfo {volume} {52}},\
  \bibinfo {pages} {295} (\bibinfo {year} {1937})}\BibitemShut {NoStop}%
\bibitem [{\citenamefont {Hansen}(2013)}]{Hansen2013}%
  \BibitemOpen
  \bibfield  {author} {\bibinfo {author} {\bibfnamefont {K.}~\bibnamefont
  {Hansen}},\ }\href@noop {} {\emph {\bibinfo {title} {Statistical physics of
  nanoparticles in the gas phase}}},\ \bibinfo {series} {Springer Series on
  Atomic, Optical, and Plasma Physics}, Vol.~\bibinfo {volume} {73}\ (\bibinfo
  {publisher} {Springer Dordrecht},\ \bibinfo {year} {2013})\BibitemShut
  {NoStop}%
\bibitem [{\citenamefont {Beyer}\ and\ \citenamefont
  {Swinehart}(1973)}]{Beyer1973}%
  \BibitemOpen
  \bibfield  {author} {\bibinfo {author} {\bibfnamefont {T.}~\bibnamefont
  {Beyer}}\ and\ \bibinfo {author} {\bibfnamefont {D.~F.}\ \bibnamefont
  {Swinehart}},\ }\href {\doibase 10.1145/362248.362275} {\bibfield  {journal}
  {\bibinfo  {journal} {Commun. ACM}\ }\textbf {\bibinfo {volume} {16}},\
  \bibinfo {pages} {379} (\bibinfo {year} {1973})}\BibitemShut {NoStop}%
\end{thebibliography}%

\end{document}